\documentclass[authoryear,preprint,12pt]{elsarticle}
 
\usepackage{amssymb}
\usepackage{multirow}
\usepackage{booktabs,subcaption,amsfonts,dcolumn}
\usepackage{multirow}
\usepackage{rotating}
\usepackage{algorithmic}
\usepackage{graphicx}
\usepackage{amsmath}
\usepackage{url}
\usepackage{float}
\usepackage{geometry}
\usepackage{color}
\usepackage{dsfont}
\usepackage[ruled,vlined]{algorithm2e}
\usepackage{hyperref}
\usepackage[x11names]{xcolor}
\usepackage{pgf, tikz} 
\usetikzlibrary{arrows,automata,fit}
\usetikzlibrary{arrows.meta}
\usetikzlibrary{shapes}
\usetikzlibrary{arrows,shapes.geometric,graphs,graphs.standard,positioning}
\usepackage{makecell}

\usepackage{bm}
\usepackage{lscape} 

\begin{document}

\begin{frontmatter}

\title{An analysis of factors impacting team strengths in the Australian Football League using time-variant Bradley-Terry models}

\author{Carlos Rafael Gonz\'{a}lez Soffner}
\author{Manuele Leonelli}

\affiliation{organization={School of Science and Technology, IE University},
            city={Madrid},
            country={Spain}}

\begin{abstract}
Australian Rules Football is a field invasion game where two teams attempt to score the highest points to win. Complex machine learning algorithms have been developed to predict match outcomes post-game, but their lack of interpretability hampers an understanding of the factors that affect a team’s performance. Using data from the male competition of the Australian Football League, seasons 2015 to 2023, we estimate team strengths and the factors impacting them by fitting flexible Bradley-Terry models. We successfully identify teams significantly stronger or weaker than the average, with stronger teams placing higher in the previous seasons’ ladder and leading the activity in the Forward 50 zone, goal shots and scoring over their opponents. Playing at home is confirmed to create an advantage regardless of team strengths. The ability of the model to predict game results in advance is tested, with models accounting for team-specific, time-variant features predicting up to 71.5\% of outcomes. Therefore, our approach can provide an interpretable understanding of team strengths and competitive game predictions, making it optimal for data-driven strategies and training.

\end{abstract}









\end{frontmatter}

\section{Introduction}

Australian Rules Football (AF) is a field invasion game played by two teams of 18 players each through four consecutive twenty-minute quarters. Teams score by sending the ball (football or footy) in between the opposite team’s four goalposts: teams score a goal (worth 6 points) if the ball is kicked through the central posts, and a behind (worth 1 point) if it goes through a centre and a side post, touches the post or is touched by an opponent. The team with the highest score by the end of the fourth quarter wins the game, with draws if both teams have the same score.

Players can move the football by running and bouncing it or kicking and handballing (punching the ball towards a teammate) to dispose of it. Opponents attempt to seize the ball by interrupting disposals, tackling (forcing the player holding the ball to dispose of it) or taking a mark (catching a football that was kicked in the air, awarding a free kick). The playing ground has an oval shape and can be divided into three zones: Defensive (50 metres closest to a team’s goalposts), Forward 50 (50 metres closest to the opponent’s goalposts) and Midfield (area in between).

The male Australian Football League (AFL) is the prime AF competition, gathering 18 teams across Australia. They compete against each other every week in a Home-and-Away season, usually made up of 23 or 24 rounds, and are ranked on a ladder based on their game outcomes: wins award 4 points, draws 2 points and losses no points.
The top 8 teams in the ladder progress to a Finals Series, where losers are progressively eliminated until the Grand Final. The winner of the Grand Final is referred to as the season Premier.

There is extensive research about the use of data collected during AF games to model outcomes, players’ performance or network effects, most of them using data from AFL games as these are publicly available on different websites. Researchers have successfully predicted game winners and score margins using counts of actions like handballs or disposals \citep{robertson2016explaining,young2019modelling}, network measures \citep{braham2018complex,fransen2022cooperative,sargent2022sleep,young2020understanding} or features of the game itself, like playing home or away \citep{robertson2018evaluating}.

Most of the studies modelling game outcomes take a post-game approach, aiming to predict the winner or the scores through their in-game performance. While often resulting in well-performing models \citep{robertson2016explaining,young2019modelling,young2020understanding}, any derived conclusion cannot give information about a team’s future performance beyond what went wrong during a game or what strategies worked best. On the other hand, research efforts to model outcomes pre-game, while achieving worse-performing models \citep{fahey2019multifactorial,manderson2018dynamic,sargent2022sleep}, do confirm the viability of this approach, whereby teams’ performance in their past games can be predictive of their future performance. 

Additionally, even though complex machine learning models have performed best \citep{fahey2019multifactorial,manderson2018dynamic,young2019modelling,young2020understanding}, in the context of sports they can be problematic since their interpretation must be comprehensible enough for teams to modify their training and strategies accordingly. Thus, there is a tradeoff between advanced predictive methods and the need to  understand what makes a team more likely to win a game before it takes place.

We explore the performance of various, flexible Bradley-Terry (BT) models \citep[see e.g.][]{agresti2012categorical} to estimate teams' strength and predict AF games' outcomes using pre-game features. The BT model is designed to model pairwise comparisons, which are extremely common in sports like AF since each team is pitted against another. For a game between two hypothetical teams $i$ and $j$, the model calculates the probability of team $i$ winning as a measure of the strength of team $i$ relative to the combined strength of teams $i$ and $j$. While the original model formulation assumes that the values for the strengths are fixed in time, \citet{tsokos2019modeling} describes expansions of the model to include time-variant features. Moreover, the \texttt{BradleyTerry2} R package further adapts the model to game and team-specific features that can vary with time \citep{firth2012bradley}.
This is the first paper to attempt using the BT model to predict outcomes in AF, despite its satisfactory performance in other sports, such as football \citep{tsokos2019modeling}, cricket \citep{dewart2019using}, and tennis \citep{fayomi2022forecasting}.

Our aim is to determine how well various BT models can predict AF match outcomes with features only available before the game takes place. The games predicted are those from the male competition of the AFL, seasons 2015 to 2023,  as the majority of the research efforts to date have concentrated on this competition. 
 BT models of increasing complexity are fitted progressively, from its simplest specification, which only considers binomial counts of outcomes, to the most complex, which predicts games with time-variant predictors. Our approach deviates from others in its encoding of some of the features to account for a team’s cumulative performance over a whole season or the last 4 games. Furthermore, we fit models over different windows of seasons to investigate whether a higher sample size of past seasons improves the fit and predictive ability.

The paper is organised as follows. Section \ref{sec:related} reviews the state of the art in the features, models and techniques used to predict AFL game outcomes. Section \ref{sec:methodology} describes the data collection and feature engineering processes, the BT models 
and the experiment design. The significant features and models’ performance are presented in Section \ref{sec:results}, along with a discussion in Section \ref{sec:discussion} on the implications for strength estimation in the AFL. Section \ref{sec:conclusions} concludes the paper.

\section{Related work}
\label{sec:related}


\subsection{Game features}

The different movement and scoring possibilities in AF, along with the complex fixture design of the AFL, result in a broad pool of possible features to model game outcomes. Most research efforts have concentrated on variables relating to the fixture (e.g.: location of the game, playing home or away) or the team totals of actions of individual players (e.g.: total kicks, total goals), although there is a growing body of research about team-specific features represented as network effects.

Following \citet{fahey2019multifactorial}, this section identifies ‘strength’ and ‘match difficulty’ features. The former refers to the variables potentially identifying that a team is stronger than another regardless of fixed effects like the location of the game. The latter covers the features impacting the odds of winning regardless of how strong a team is compared to their opponent (e.g.: playing away, travelling).

\subsubsection{Technical strength}

Technical features are the individual and aggregated actions of the players and the results of the teams. The most prolific ones in the literature are Performance Indicators (PIs), the actions individual players perform during a game that can be aggregated into team totals. Modelling game outcomes based on the absolute counts of the indicators (say, total kicks of a team during a game) is discouraged as it dismisses the performance of the opponent \citep{robertson2016explaining}. Indeed, \citet{young2019modelling} confirmed that PIs in their relative form were significantly more explanatory of game outcomes and score margins than PIs in their absolute form.

Achieving higher relative counts of kicks and goal conversions during the game appears to be indicative of winning. More specifically, during the 2013 season, 81\% of the teams recording kick differentials higher or equal to $-1$ won their games, while 90\% of the teams with higher additional kicks and a goal conversion 4.2\% higher than their opponents also won their games \citep{robertson2016explaining}.  Furthermore, it is natural that teams with more activity in the Forward 50, the zone closest to the opponent’s goalposts, are more likely to score. Indeed, \citet{robertson2016explaining} mentioned that differentials in the inside 50s (entries into the Forward 50), marks in the Forward 50, regular marks and contested possessions were all influential variables for in-game outcomes. The findings of \citet{young2019modelling} confirmed the effect of all previous indicators, along with the differentials in disposals, time in possession and metres gained, the last two being more important than any other indicator.

Notably, most research about PIs is descriptive, with the outcomes or score margins modelled with the features observed in such a game. While still relevant for post-game analyses, there is a research gap on the predictive ability of cumulative PIs before the game takes place. This is particularly striking considering that the evolution of these indicators over the season or a window of past games is the simplest indicator of a team’s relative strength to another.

In terms of the individual players' strength, it has been found that older and heavier teams, that is, with a higher average player age and weight, are significantly (yet slightly) more likely to win their games \citep{lazarus2018factors}. It is theorised that older players’ experience may compensate for any physical deterioration.

While most research efforts have concentrated on understanding a team’s strength as the aggregate of their players, some studies have focused on team-level strength indicators. One of these indicators is the team's performance over past games. The wins over a window of the past 4 games, known as ‘momentum’ or ‘form’, were found to significantly increase the likelihood of winning a game \citep{robertson2018evaluating}. Furthermore, the difference in the ladder position was also associated with a slightly higher chance of winning \citep{robertson2018evaluating}, along with differentials in the ladder rank from the previous season \citep{fahey2019multifactorial,robertson2018evaluating}.

Finally, differentials in higher rankings in player-based team ratings, Elo ratings (ratings of players or teams in two-player games, updated based on wins and losses and the rating of the opponent) and the availability of the Top 10 best-performing players were significant predictors of game outcomes  \citep{fahey2019multifactorial}. However, while being calculated before the game takes place, except for the Elo team ratings, there is no consensus over a standardised way to rank players that all teams can use.

\subsubsection{Tactical strength}

Predictions of game outcomes based on the aggregation of individual PIs have lately been reexamined in favour of approaches modelling AF as Social Networks built over the players’ interactions, often referred to as a ‘tactical’ analysis. 

In one of the earliest tactical analyses of AF, \citet{sargent2013evaluating} modelled games as passing networks, with edges representing passes between players, and concluded that higher values for a team’s connectedness were predictive of the score margin, to the point that they could estimate the impact of important players missing and recommend potential substitutes. \citet{sargent2013evaluating} therefore suggests that it is possible to predict the outcome of a game once the lineup is revealed. \citet{braham2018complex} built similar ‘passing networks’, confirming that stronger teams tend to show higher network values. More specifically, the average values for a team’s betweenness and out-degree were important predictors of game outcomes, along with the mean betweenness for the previous game and the previous four games \citep{braham2018complex}. 

\citet{young2019understanding}, however, found no significant differences in the betweenness centrality between winning and losing teams. While the other network measures remained significant, they were slightly to moderately correlated with the score margin. It could be possible that network measures might explain playing strategies that remain stable throughout a season but change from season to season, perhaps to compensate for bad results or reflect changes in coaching strategies or adaptations to rules. Furthermore, just like with PIs, absolute values of network measures may not fully capture their effect, if any, on game outcomes.

Modelling approaches combining tactical and technical features are currently limited and inconclusive. Reasons for this include the difficulty of finding network data from free sources and the significant effort to calculate network measures for every team at every game. Moreover, the variety of possible actions in AF opens the door for different network designs, with the resulting network measures representing different aspects of the game. Consequently, the same measures can have different significance in different models. Indeed, \citet{young2020understanding} drew mixed conclusions from a combined approach, stating that tactical features were slightly more important than technical PIs while not improving the classification accuracy of winning and losing teams. 

However, after encoding a passing network’s interactions in three types of metrics (receiving the football as indegree, distributing as outdegree and mutual passes as connectedness), \citet{fransen2022cooperative} found that one-unit increases in the connectedness increased the probability of winning the game by 5.3\%, further arguing that tactical features explained 14\% to 27\% of the variance in the result and that their effects were stable across seasons. Consequently, the research around the inclusion of tactical features to model game outcomes might be at a stage too early to decide on a specific network design and encoding strategy.

\subsubsection{Match difficulty}

Match difficulty features are pre-game fixed effects impacting the winning probability for all teams regardless of their strengths. Accounting for these factors is important in AF, as teams sometimes travel long distances with changing rest days between games \citep{sargent2022sleep}. 

One significant condition is the Home and Away team designations. It is common for teams to play games in shared home venues (say, North Melbourne and Saint Kilda share the Marvel Stadium) or grounds with no teams in the AFL (mostly in the Northern Territory and Tasmania). \citet{robertson2016explaining} found that teams playing away were less likely to win the game, especially if they had travelled to another state, an effect that \citet{lazarus2018factors} confirmed to be significant yet small. The effect of playing at home or as the Home Team has been explored widely in other sports, with some researchers pointing to the positive effects of emotional support from fans \citep{leitner2023cauldron}.

Regarding the travelling effect, it is also possible that players rest less, especially considering the different time zones in Australia and the impact it can have on physical performance. While it has been confirmed that playing interstate games decreases the sleep of players, there are no studies analysing the effect of poor sleep on game outcomes \citep{sargent2022sleep}.

\subsection{Eras in AFL}

Like any long-running sport, AF has experienced changes in its rules and style of play. The current game is faster, with heavier and taller players \citep{woods2017evolution}. Scoring has also declined in the latest seasons \citep{lane2020characterisation}. While PIs and strategies remain stable throughout the same seasons, a growing body of research argues that changes to the game can be appreciated in the medium term, to the point of identifying groups of seasons or ‘eras’ in the AFL where the indicators are significantly different from each other.

\citet{woods2017evolution} confirmed how the totals of PIs varied in different seasons, possibly reflecting changing strategies or adaptation to new rules of the game. More specifically, they identified an era where possessing the football was prioritised (2005-09 seasons), followed by a defensive era (2010-13), and an era focusing on repossession and turnovers (2014-onwards). Likewise, \citet{lane2020characterisation} identified an era where offensive strategies predominated (until 2007-08), followed by an increasingly defensive game (2007-11) with more tackles and disposals and lower scoring.

Identifying eras raises doubts about the effectiveness of a historical model making no distinction between seasons. If PIs change between eras, modelling all seasons could imply trying to fit one model to different strategies of the same game. Similar concerns were voiced by \citet{young2019modelling}, as the lack of proper consensus on the methods for identifying these periods complicated the modelling and validation processes. Indeed, while there is not a unique strategy for identifying eras, the findings do warrant a model that is either season-specific or assigns heavier weights to the most recent seasons or rounds.

\subsection{Machine learning models}

Research shows that non-linear models often perform better when modelling outcomes in the AFL, especially if they perform feature selection \citep{fahey2019multifactorial}. Such models are common in team invasion sports and feature selection is necessary before or during modelling given the high number of possible predictors. In the context of pre-game outcome predictions, black-box and complex machine learning models are not appropriate as their predictive power does not compensate for the lack of interpretability. 

Research often centres on classification models to predict binary outcomes (wins or losses), excluding draws since they are rare in AF. However, the scoring dynamics of the sport have drawn attention to models predicting score margins. Strangely enough, research about modelling the number of goals or behinds, jointly or separately, is sparse.

Models based on decision trees often yield good results, which follows the notion that values above a threshold in differential PIs can successfully identify winning teams. \citet{young2020understanding} used a decision tree to model wins and losses with technical and tactical indicators, achieving a classification accuracy of about 89\%. Similarly, random forest models achieved accuracies between 85\% and 89\% when modelling outcomes in different eras \citep{young2019modelling}.

Generalised linear models (GLMs) often outperform other techniques. \citet{young2020understanding} confirmed that GLMs achieved a classification accuracy of  93.2\% when modelling wins and losses, and an RMSE of 6.9 points when modelling the score margin. GLMs have also been used to predict outcomes in advance. Specifically, training a GLM with elastic net regularisation on features known before the game, \citet{fahey2019multifactorial}  reported an accuracy of 73.3\% on the 2018 season games and 71.6\% during cross-validation (seasons 2013-17).

Finally, \citet{manderson2018dynamic}, arguing that the number of goals and the number of behinds each followed a Skellam distribution, used a Bayesian hierarchical model to produce pre-game predictions on the two scoring types, identifying the team with the highest aggregated points as the winner. They correctly predicted 132 games (around two-thirds), performing slightly worse than betting markets.

All the literature on machine learning approaches for AF reviewed above is further summarized in \ref{appendixa}.

\section{Methodology}
\label{sec:methodology}
\subsection{The data}

Data was fetched through the \texttt{fitzRoy} R package, which offers different functions to fetch official AFL data from various sources, including the AFL, AFL Tables, FootyWire and Squiggle \citep{nguyen2021fitzroy}. It is currently the most comprehensive, non-commercial tool to access AFL data and has been used in previous research \citep{fahey2019multifactorial}. Match results from seasons 2015 to 2023 were retrieved from AFL Tables through \texttt{fitzRoy}. Ladder results from the same seasons were retrieved from the AFL official website, also through the package. The ladder result from the last round of the Home-and-Away season, combined with the placement at the Finals Series, was retrieved manually from the AFL Website. Finally, \texttt{fitzRoy} was used to retrieve performance indicator data for each game in those seasons from the AFL Website. Network data was not retrieved as it was not publicly available, thereby discarding the use of tactical features.

Data from 1826 games from seasons 2015 to 2023 was used to fit the models. These amounted to 207 games per season except for 2015 (206 as one game was cancelled), 2020 (162 since the season was cut short due to Covid) and 2023 (216 games due to the addition of one round). Following \citet{fahey2019multifactorial} and \citet{fransen2022cooperative}, draws were excluded as they amounted to only 14 games. A glossary with the details of all features used to predict games' outcomes is given in \ref{glossary}, while \ref{prep} summarizes the data pipelines used to create the required dataframes to fit the models.

The code to extract the data and replicate the experiments can be found at \href{https://github.com/charlieceratops/AFL\_BradleyTerry}{https://github.com/charlieceratops/AFL\_BradleyTerry}.

\subsection{The Bradley-Terry model}

The Bradley-Terry (BT) model is a GLM used to model pairwise comparisons, like those in AF, where two teams compete against each other for the win. Following the explanation in \citet{tsokos2019modeling}, AF outcomes can be modelled like:

\[
Y_{ijr} = \left\{\begin{array}{lc}
1, & \mbox{if team } i \mbox{ beats team } j \mbox{ at round } r,\\
0, & \mbox{if team } j \mbox{ beats team } i \mbox{ at round } r,\\
\end{array}\right.
\]
for $i,j=1,\dots,18$, $i\neq j$, since there are 18 teams in the AFL, and $r=1,\dots,N$, where $N$ is the total number of rounds. The BT model takes the probability of Team $i$ winning against Team $j$ in round $r$ as:
\[
P(Y_{ijr} = 1) = \frac{\pi_{ir}}{\pi_{ir}+\pi_{jr}},
\]
where $\pi_{ir} = \exp(\lambda_{ir})$ and $\lambda_{ir}$ is the strength of Team $i$ in round $r$ \citep{tsokos2019modeling}. This formulation simplifies the understanding of a game outcome as proportional to how strong a team compares to the other, thereby leading to more understandable explanations of team performance. 

\subsubsection{Standard Bradley-Terry}
The initial formulation of the BT model assumes strengths and features to be time-invariant \citep{tsokos2019modeling}. 
Teams’ strengths to depend only on game outcomes, excluding any contest or team-specific predictors. Assuming independence of games, the standard BT models a binomial target (Team $i$, Team $j$) based on each team’s wins for each pairing \citep{firth2012bradley}. The resulting coefficients are the logarithms of the strengths of each team. The probability of Team $i$ winning in round $r$ against Team $j$ is therefore calculated as
\[
\textnormal{logit}(P(Y_{ijr}=1)) = \lambda_{i}-\lambda_{j},
\]
where we dropped the $r$ subscript since strengths are constant throughout a season.

\subsubsection{Contest-specific, time-invariant Bradley-Terry}

The contest-specific, time-invariant BT model expands the standard one to allow for ‘order effects’, which are biases in favour of a particular team, thereby extending the model to
\[
\textnormal{logit}(P(Y_{ijr}=1)) = \lambda_{i}-\lambda_{j} + \delta Z
\]
where $\delta$ is the value of the bias and $Z = 1$ if $i$ has the advantage, $-1$ otherwise \citep{firth2012bradley}.

In the AFL, the only effect that is specific to all contest pairs and constant over time is whether or not the team is designated as Home or Away. Other possible features will be specific to the teams or time-variant (PIs and past performance). Moreover, since teams can be designated as Home in games outside their home venue and state, it is not possible to derive order effects from these features.

The model estimates the order effect along with the team strengths, which are independent of it. Conceptually, team strengths are still derived from the frequency of wins since there are no other team-specific features.

\subsubsection{Team-specific, time-variant Bradley-Terry}

This final expansion allows the introduction of features with values specific to the competing teams (e.g.: the number of wins through the season), with the strength of the team being calculated as

\[
\lambda_{ir} = \sum_{k=1}^p\beta_kX_{ik}+\varepsilon_{i},
\]
where $\varepsilon_i\sim N(0,\sigma^2)$ are Gaussian independent errors, $X_{i1},\dots,X_{ip}$ are independent features to predict Team $i$ strength, and $\beta_1,\dots,\beta_p$ are the features coefficients.

Following the literature review, time-invariant features will not be included in the modelling. This is because most of the strength features, as described below, are taken as differentials, meaning their value changes for each game relative to the opposite team. Moreover, measures of game difficulty beyond the AT\_HOME effect are all player-specific and time-variant.  Consequently, because team strength is dependent on player-specific features that change for each game, the BT model will no longer return fixed estimates of team abilities, as these are not explained by the binomial counts of outcomes but by predictors.

\subsection{Construction of features}
The ladder position, that is the position after the previous game, is firstly added to the game results. 
The value is transformed into differentials following \citet{robertson2018evaluating} as it is more predictive of game outcomes compared to the absolute value. As in \citet{fahey2019multifactorial} and \citet{robertson2018evaluating}, the differentials of the ladder position of the last year are also calculated for both teams.

Team form features are used to assess short and long-term performance throughout the season. These include whether a team won their last game, the number of games won over the last 4 games, the cumulative wins over the season and the consecutive wins and losses. Only the differential from the cumulative wins is taken.

Differentials for the points in favour, points against and the percentage ratio of the first variable to the latter are calculated as the absolute difference. PIs are aggregated for each game and team, summed to be taken as cumulatives over the season and the last 4 games, and calculated as differentials. Essentially, the past performance of two competing teams is compared.

Finally, variables to assess match difficulty are calculated, particularly binary variables assessing whether the team is playing at home, at their home ground or if they have travelled interstate. 

\subsection{Description of the experiments}

\subsubsection{Experiment 1: Standard Bradley-Terry}

The \texttt{BradleyTerry2} R package was used to fit the standard BT specification on each season and an additional dataset containing all games from seasons 2015 to 2023. The model was further trained on windows of 2, 3 and 4 seasons to explore cross-seasonal variations in team strengths. The target variable (WIN) indicated whether the team won the game (1) or not (0), excluding draws as explained before. The goodness of fit was assessed with the Akaike Information Criterion (AIC). To simplify the interpretation of the results, for each window of seasons, the reference team/level was chosen as the team which had an 'average performance' (half wins/half losses).

\begin{figure}[H]
    \centering
    \includegraphics[scale=0.6]{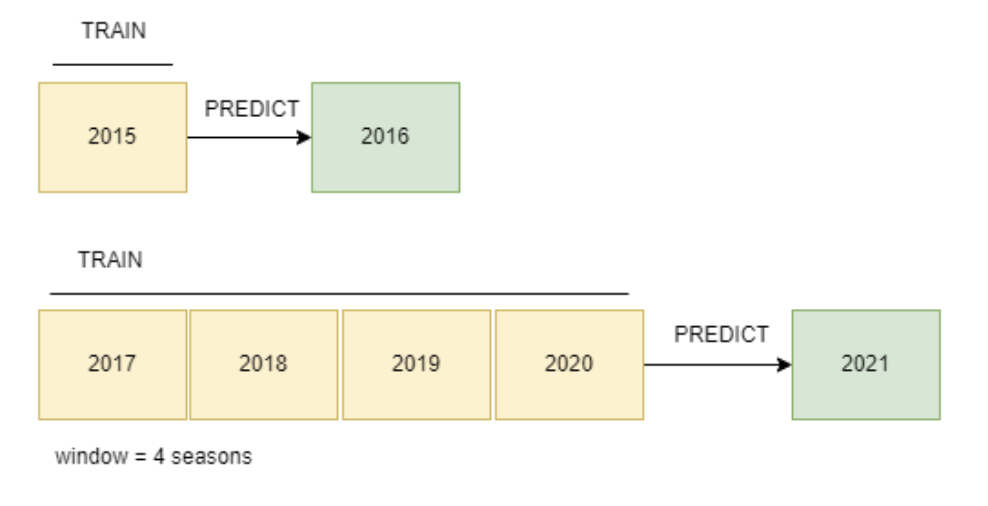}
    \caption{Examples of training and testing of models in Experiment 1. In the first example, a model is trained on 2015 results and predicts all the games in 2016. In the example below, a model is trained on a window of 4 seasons, from 2017 to 2020, to predict all the games in 2021.}
    \label{fig:description1}
\end{figure}

Classification accuracy was calculated for each model on the predicted probability of the Home Team winning in the training dataset and the test dataset for the following season. For the models trained on more than one season, the predicted season was the following to the last season in the training data. A team was predicted to win whenever their probability was higher than 0.5. Figure \ref{fig:description1}. summarises the training and testing strategy for this experiment.

\label{sec:exp1}

\subsubsection{Experiment 2: Contest-specific effects}

This experiment makes no changes to the model training and testing explained in Section \ref{sec:exp1}, with the difference of the added AT\_HOME effect, estimated as a parameter separate from the team strengths.

\subsubsection{Experiment 3: Team-specific, time-variant features}

In this experiment, models are trained on games for each season and used to predict all the games in the following, using windows of one or two seasons. The target variable remains the same and the predictive ability is estimated with the classification accuracy.

The features used for training can be divided into three categories: Match Difficulty (e.g.: playing home or away), PIs and Form (measures teams’ performance e.g.: wins over last 4 games). Models are further trained on two sets of features: one with the cumulative performance over the season, and the other with the cumulative performance over the last four games. 

\begin{figure}[H]
    \centering
    \includegraphics[scale=0.56]{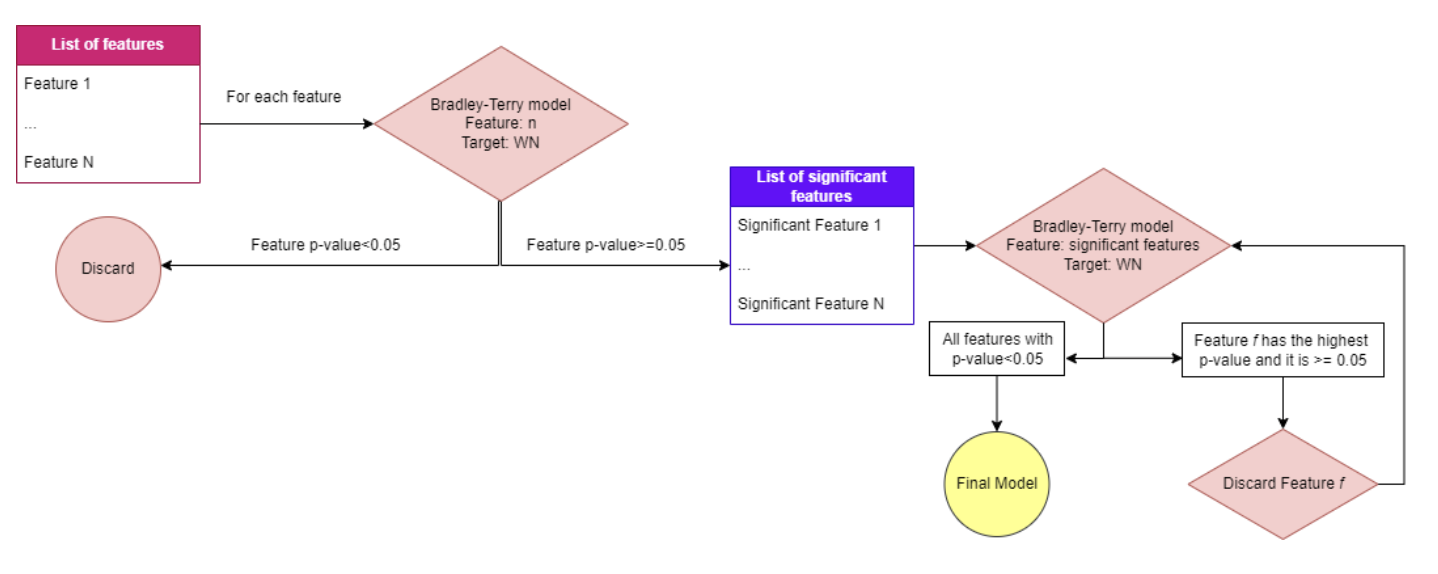}
    \caption{Process for training models in Experiment 3. }
    \label{fig:description2}
\end{figure}

Following the recommendations in the literature to perform feature selection \citep{fahey2019multifactorial}, a model is trained on one season or window of seasons for each individual feature which, if found to be significant at the 5\% level, is added to a set of final features. A final model is trained on the significant features, performing backward elimination of the feature that has the highest p-value above 0.05 until all the remaining features are significant. The features included in the final model are recorded along with the features that were found to be significant individually, although these are not part of the model used for predictions. Figure \ref{fig:description2} illustrates the training process.

\subsubsection{Experiment 4: Round-by-round predictions}
\label{sec:strategy}

This experiment expands on the previous models, changing the prediction strategy. The reason behind it is that, in a real scenario, models can be retrained throughout the season with updated information, instead of predicting the full season in advance with historical data. We now predict games in the test season, updating the models throughout the testing process, which is the closest simulation of a BT model that would allow for time-varying coefficients. Predictions are made round-by-round instead of game-by-game since it is common for AFL games to overlap each other. 

We test four different strategies to perform round-by-round predictions. The \textit{Addition} strategy trains a model on a single season and tests the model on the first round of the following season. Once the games are predicted, data from that round is added to the training data and the model is retrained. The model is tested and retrained for each round until the Grand Final. The \textit{Substitution} strategy is similar, although the data from the corresponding round of the previous season is discarded during the retrain to give more weight to the current season.

\begin{figure}[H]
    \centering
    \includegraphics[scale=0.6]{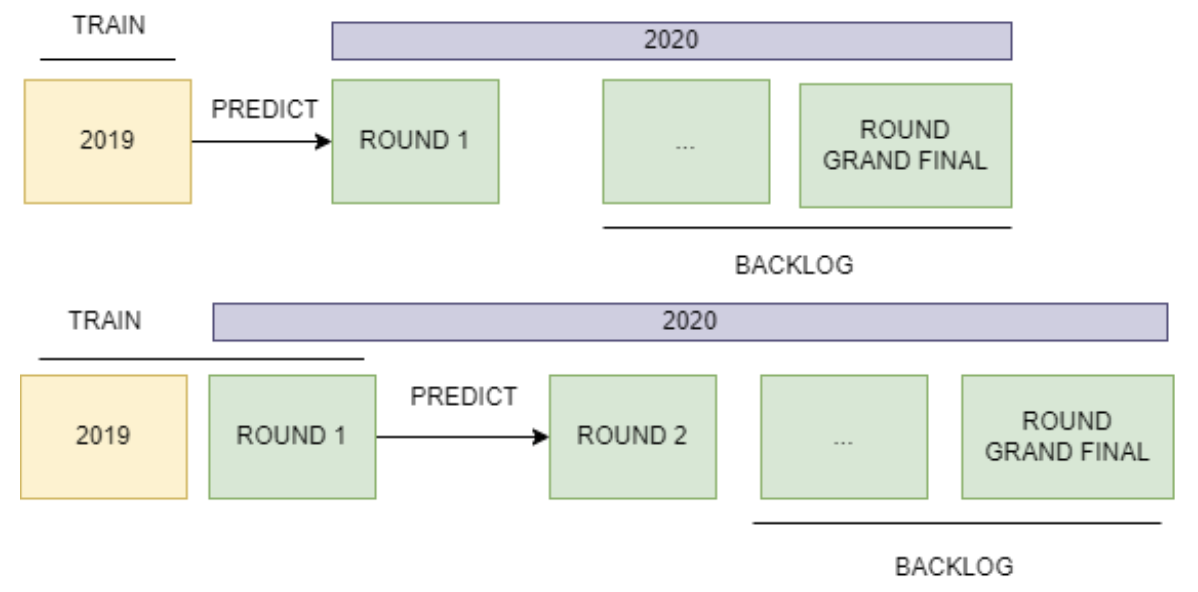}
    \caption{Example of the Addition strategy. The model trained on 2019 games is used to predict Round 1 of the 2020 season. Then, the model is retrained with that season and used to predict Round 2. The process continues until the Grand Final.}
    \label{fig:description3}
\end{figure}

\begin{figure}[H]
    \centering
    \includegraphics[scale=0.6]{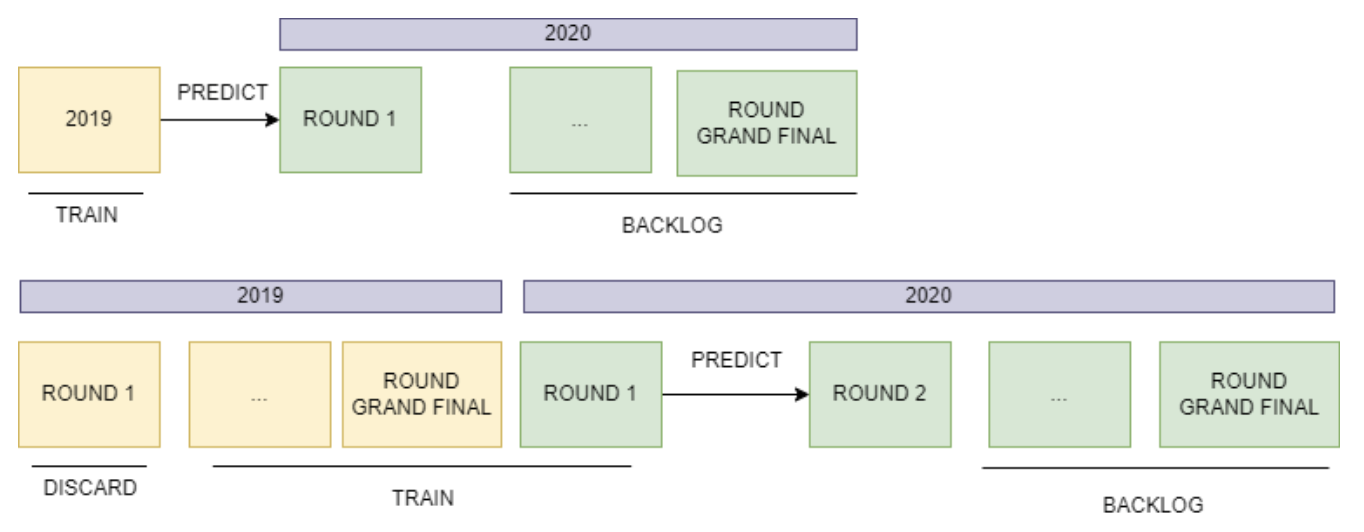}
    \caption{Example of the Substitution strategy. The model trained on 2019 games is used to predict Round 1 of the 2020 season. Then, the first round of 2019 is discarded, and the model is retrained with the remaining rounds and the first round of 2020 and used to predict Round 2. The process continues until the Grand Final.}
    \label{fig:description4}
\end{figure}

\begin{figure}[H]
    \centering
    \includegraphics[scale=0.6]{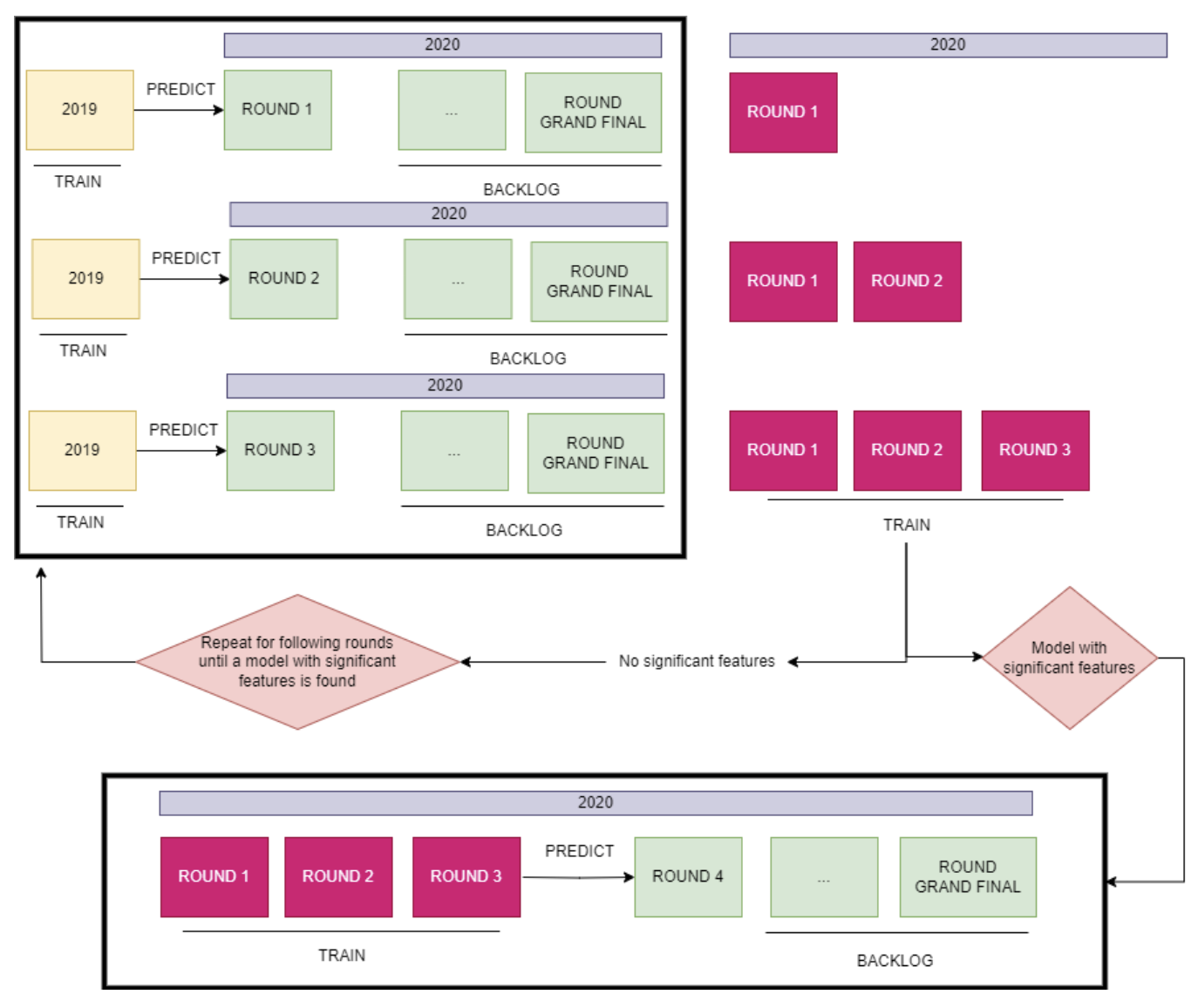}
    \caption{Example of the Incremental strategy. The model trained on 2019 games is used to predict the first three rounds of the 2020 season. If the model fitted to those rounds has no significant features, the process is repeated for the following round. Otherwise, the newly trained model is used to predict the subsequent rounds.}
    \label{fig:description5}
\end{figure}

In the \textit{Incremental} strategy, the first three rounds of the test data are predicted with the model trained on the previous season since most of the differential features will amount to 0.
From the third round onwards, a model is built on the tested rounds: if no significant features are found, the predictions will be those from the previous season model; otherwise, the model with significant features will be used to predict the subsequent rounds. Figures \ref{fig:description3}-\ref{fig:description5} illustrate the three strategies.

Finally, the \textit{Majority Voting} strategy combines the predictions from these three strategies along with the models in Experiments 2 and 3, making a final prediction by majority voting. A team is predicted to win if at least three of the predictions from the chosen models predict it to win.


\section{Results}
\label{sec:results}

\subsection{Team Strengths Based on Game Outcomes (Experiment 1)}

The estimated strength coefficients for the standard BT model trained on windows of one season are summarised in Table \ref{tab:coefficients}. TABLES \ref{tab:coefficientsw2} to \ref{tab:coefficientsw4}. display the coefficients for the models trained on windows of two, three and four seasons. 

\begin{table}
    \centering
    \caption{Estimated strength coefficients with the standard Bradley-Terry model, fitted to single seasons (2015-2023) and all available data. Coefficients found to be significant at the 5\% level are in bold and underlined. The coefficient for the reference team is 0.00.}
    \label{tab:coefficients}
    \scalebox{0.6}{
    \begin{tabular}{l *{10}{c}}
    \toprule
    \multicolumn{1}{c}{\textbf{Team}} & \multicolumn{10}{c}{\textbf{Season}} \\
    \cmidrule(lr){2-11}
    & 2015 & 2016 & 2017 & 2018 & 2019 & 2020 & 2021 & 2022 & 2023 & 2015-23 \\
    \midrule
    Adelaide & 0.14 & \textbf{\underline{1.55}} & 1.15 & 0.04 & -0.29 & -1.58 & -0.77 & -1.07 & 0.00 & 0.00 \\
    Brisbane Lions & \textbf{\underline{-2.34}} & \textbf{\underline{-2.08}} & -1.27 & \textbf{\underline{-1.74}} & 0.72 & \textbf{\underline{1.80}} & 0.75 & 0.78 & \textbf{\underline{1.31}} & -0.09 \\
    Carlton & \textbf{\underline{-2.26}} & -0.94 & -1.00 & \textbf{\underline{-3.03}} & -0.99 & -0.28 & -0.56 & 0.04 & 0.61 & \textbf{\underline{-0.63}} \\
    Collingwood & -0.79 & -0.21 & -0.26 & 0.66 & 0.81 & 0.43 & -0.95 & 0.82 & \textbf{\underline{1.75}} & 0.25 \\
    Essendon & \textbf{\underline{-1.57}} & \textbf{\underline{-2.12}} & 0.06 & 0.10 & 0.09 & -0.40 & -0.06 & -0.95 & -0.13 & -0.35 \\
    Fremantle & 1.03 & -1.55 & -0.53 & -0.91 & -0.43 & -0.28 & -0.13 & 0.76 & -0.36 & -0.16 \\
    Gold Coast & \textbf{\underline{-2.20}} & -1.35 & -1.17 & \textbf{\underline{-2.27}} & \textbf{\underline{-2.02}} & -0.74 & -0.76 & -0.58 & -0.48 & \textbf{\underline{-0.95}} \\
    Geelong & 0.00 & \textbf{\underline{1.86}} & 0.90 & 0.20 & 0.84 & 1.22 & 1.14 & \textbf{\underline{1.68}} & -0.10 & \textbf{\underline{0.66}} \\
    Greater Western Sydney & -0.67 & \textbf{\underline{1.58}} & 0.89 & 0.40 & 0.56 & 0.00 & 0.39 & -1.20 & 0.45 & 0.24 \\
    Hawthorn & 0.92 & \textbf{\underline{1.63}} & -0.07 & 0.41 & 0.08 & -0.87 & -0.65 & -0.96 & -0.97 & 0.00 \\
    Melbourne & -1.36 & -0.01 & 0.17 & 0.42 & -1.32 & 0.28 & \textbf{\underline{2.04}} & 0.83 & 0.73 & 0.20 \\
    North Melbourne & 0.20 & 0.75 & -1.09 & -0.25 & -0.16 & -1.58 & \textbf{\underline{-1.48}} & \textbf{\underline{-2.83}} & \textbf{\underline{-2.25}} & \textbf{\underline{-0.66}} \\
    Port Adelaide & 0.02 & 0.00 & 0.46 & 0.00 & 0.00 & \textbf{\underline{1.80}} & \textbf{\underline{1.36}} & -0.36 & 0.96 & \textbf{\underline{0.39}} \\
    Richmond & 0.31 & -0.49 & 1.09 & 1.33 & 1.25 & \textbf{\underline{1.67}} & -0.13 & 0.07 & -0.21 & \textbf{\underline{0.49}} \\
    Saint Kilda & \textbf{\underline{-1.60}} & 0.44 & 0.07 & \textbf{\underline{-1.86}} & -0.68 & 0.61 & 0.00 & 0.00 & 0.15 & -0.22 \\
    Sydney & 0.50 & \textbf{\underline{1.82}} & 0.60 & 0.44 & -0.70 & -0.87 & 0.75 & 0.96 & 0.14 & 0.36 \\
    Western Bulldogs & 0.05 & \textbf{\underline{1.76}} & 0.00 & -0.93 & 0.12 & 0.43 & 1.21 & 0.05 & 0.09 & 0.26 \\
    West Coast & 1.05 & 1.38 & 0.29 & 1.15 & 0.70 & 0.96 & -0.17 & \textbf{\underline{-2.79}} & \textbf{\underline{-2.22}} & 0.14 \\
    \bottomrule
    \end{tabular}
}
\end{table}

Not all strength coefficients in Table \ref{tab:coefficients} are significant in each season, with seasons like 2017 having no teams being significantly stronger or weaker than the reference (average) team. The direction of the coefficients is also not constant: notice how the significant strengths in 2015 are all negative while those in 2020 are positive. Finally, while the window of outcomes from 2015 to 2023 does mark the strengths of teams that have “historically” performed worse (Gold Coast) or better (Geelong) as significant, it estimates the strengths of teams with an evolving performance (Brisbane Lions or West Coast) close to 0.

The significance of strength estimates varies when considering windows of more than one season. Windows of two seasons (Table \ref{tab:coefficientsw2}) will show Richmond being significantly stronger from 2017 to 2020 while, in the previous analysis, they were only significant in 2020; or Carlton, who were consistently weaker from 2015 to 2019, and not just in 2015 and 2018. Windows covering three (Table \ref{tab:coefficientsw3}) or four (Table \ref{tab:coefficientsw4}) seasons support the previous findings, confirming some teams a significantly stronger than others throughout the window (Richmond, Geelong) and capturing changing results in others (Brisbane Lions).

Table \ref{tab:aic_accuracy} shows the AIC and classification accuracy of the fitted models. As to the goodness of fit, the AIC is similar within windows, except for the 2020 season, where it decreases, while increasing for windows of more seasons. The classification accuracy over the season where the model was trained is lowest at 65.38\% for 2017 and highest at 78.38\% for 2016. The metric progressively decreases with the windows of more seasons (Table \ref{tab:aic_accuracy_standardw2w4}).

The classification accuracy over the tested seasons is lower than that in the training set, ranging around 60\% for windows of one season and 60.98\% for all available data. On average it decreases as the windows become larger (Table \ref{tab:aic_accuracy_standardw2w4}), although this changes for some seasons like 2018 (65.76\%) and 2023 (65.26\%), with the highest accuracies achieved with a window of three seasons. 

\begin{table}
    \centering
    \caption{AIC and classification accuracy for the standard Bradley-Terry model, fitted to windows of one season (2015-2023).}
    \label{tab:aic_accuracy}
        \scalebox{0.6}{
    \begin{tabular}{c c c c c}
    \toprule
    \multicolumn{5}{c}{Window: 1 season}\\
    \midrule
 Train Season & Test Season & Train AIC & Train Accuracy & Test Accuracy \\
    \cmidrule(lr){1-2} \cmidrule(lr){3-5}
    2015 & 2016 & 237.07 & 71.20\% & 61.20\% \\
    2016 & 2017 & 224.21 & 78.38\% & 57.46\% \\
    2017 & 2018 & 269.16 & 65.38\% & 63.04\% \\
    2018 & 2019 & 233.99 & 75.54\% & 60.54\% \\
    2019 & 2020 & 269.21 & 69.73\% & 63.83\% \\
    2020 & 2021 & 196.88 & 74.83\% & 60.22\% \\
    2021 & 2022 & 254.23 & 70.33\% & 58.15\% \\
    2022 & 2023 & 237.35 & 75.00\% & 61.58\% \\
    2023 & & 261.10 & 72.90\%& \\
    \bottomrule
    \end{tabular}}
\end{table}

\subsection{Effect of home team designation (Experiment 2)}

The AT\_HOME effect was estimated along with the strength coefficients of each team. The results are summarised in Table \ref{tab:at_home_coefficient} and Tables \ref{tab:coefficients_contestw2} to \ref{tab:coefficients_contestw4}. While not significant in all seasons, the AT\_HOME effect was significant and positive when considering all available data, which translates as an advantage to the Home Team under the BT model. There are no major changes to the strength estimates.
AT\_HOME becomes significant for all training sets from the windows of three seasons onwards.

Table \ref{tab:aic_accuracy_contest} shows the AIC and classification accuracy of the fitted models. The AIC slightly decreases compared to the models in Experiment 1, denoting an increase of fit, with the classification accuracy over train showing a slight increase for all available data (61.92\%, Table \ref{tab:aic_accuracy_contestw2w4}).  The classification accuracy also improves over most test seasons, reaching highs of 67.37\% for the 2023 season.

\begin{table}
    \centering
    \caption{Estimated AT\_HOME Coefficient with the Bradley-Terry expansion for contest-specific effects, fitted to single seasons (2015-2023) and all available data. Significant coefficients at the 5\% level are in bold.}
    \label{tab:at_home_coefficient}
    \scalebox{0.6}{
   \begin{tabular}{ccccccccccc}
    \toprule
    \textbf{Season} & 2015 & 2016 & 2017 & 2018 & 2019 & 2020 & 2021 & 2022 & 2023 & 2015-23 \\
    \midrule
    \textbf{Feature (AT\_HOME)} & 0.15 & \textbf{0.63} & \textbf{0.43} & 0.25 & \textbf{0.35} & 0.35 & 0.08 & \textbf{0.62} & \textbf{0.37} & \textbf{0.29} \\
    \bottomrule
    \end{tabular}}
\end{table}

\begin{table}
    \centering
    \caption{AIC and classification accuracy for the Bradley-Terry expansion for contest-specific effects, fitted to windows of one season (2015-2023).}
    \label{tab:aic_accuracy_contest}
        \scalebox{0.6}{
    \begin{tabular}{c c c c c}
    \toprule
    \multicolumn{5}{c}{Window: 1 season}\\
    \midrule
 Train Season & Test Season & Train AIC & Train Accuracy & Test Accuracy \\
    \cmidrule(lr){1-2} \cmidrule(lr){3-5}
    2015 & 2016 & 238.33 & 72.28\% & 63.93\% \\
    2016 & 2017 & 214.80 & 75.68\% & 62.43\% \\
    2017 & 2018 & 263.75 & 65.93\% & 62.50\% \\
    2018 & 2019 & 233.93 & 77.17\% & 59.46\% \\
    2019 & 2020 & 266.42 & 68.65\% & 63.12\% \\
    2020 & 2021 & 195.64 & 76.22\% & 60.77\% \\
    2021 & 2022 & 255.98 & 71.98\% & 58.70\% \\
    2022 & 2023 & 227.32 & 78.80\% & 67.37\% \\
    2023 & & 257.97 & 72.43\% \\
    \bottomrule
    \end{tabular}}
\end{table}

\subsection{Effect of team-specific, time-variant features (Experiment 3)}

The features from the trained team-specific, time-variant BT expansions, along with those features that were found to be individually significant, yet excluded from the final models, are summarised in Table \ref{tab:bt_expansion_coefficients} and  Tables \ref{tab:bt_expansion_coefficientsseasonw1} to \ref{tab:bt_expansion_coefficientsseasonw2}.

Match difficulty features are significant in most seasons at the individual level, with at least one of them being included in the model except in 2018 and 2021. INTERSTATE and AT\_HOME are included in the model trained in all available data with strong negative and positive coefficients respectively, suggesting that teams playing interstate or away are at a disadvantage. HOMEGROUND is not included in this model, although it appears to be individually significant here and for individual seasons with a strong positive value, and it is included in four of the models in windows of two seasons. The coefficients remain similar in the models using cumulative features from the last four games and the full season. Interaction effects between AT\_HOME and the other difficulty features were tested although they were not found to be significant except for season 2015, and were thereby excluded from the following experiments.

The coefficients of the features of teams’ form do not change significantly between the models using PIs of the last four games or the season. The differential in the ladder position at the time of the game (LADDER\_POSITION\_DIFF) appears to be individually significant across seasons with a positive coefficient, although it is never included in the model. Moreover, the differentials in the previous season’s final ladder position (LADDERLY\_POSITION\_DIFF) are included in more of the final models with a positive estimate, especially those trained on individual seasons with cumulatives from the previous four games. The indicator is often included in the models trained in windows of two seasons but only until 2020-21.

The differentials in the percentage (PERCENTAGE\_DIFF) are often significant but not included in the final models. Differentials in the points in favour (POINTSFOR\_DIFF) and points against (POINTSAGAINST\_DIFF) appear in models trained in windows of one and two seasons and both sets of cumulatives, also being included in the model trained on all data with a positive coefficient. 

Whether or not a team won their previous game (LG\_WON) was included in the 2023 model with a strong, positive coefficient, and it was individually significant in the model trained on all seasons, although it was not significant for the rest of the seasons and windows. The cumulative number of wins (WINS\_CUMULATIVE\_DIFF) is significant at the individual level with a positive coefficient for some seasons, although it is included in some final models with a negative coefficient.

As to the performance indicators, most of them are significant at the individual level when considering all available data regardless of the window and encoding.

When considering PIs in their cumulative form over the previous four games, there are no major changes to their significance between windows. The differentials in the entries into the Forward 50 (INSIDE50\_L4\_CSUM\_DIFF) are only included in some of the models of the single-season windows with a positive coefficient, while being individually significant consistently when considering windows of two seasons. Likewise, differentials in goal shots (GOALS\_SHOTS\_L4\_CSUM\_DIFF) are significant with a positive coefficient between 2018-20, being included in the 2019 model, and this is further confirmed in the windows of two seasons.

Differentials in the contested possessions won at ground level over the previous four games (GETS\_GROUNDBALL50\_L\-4\_CSUM\_DIFF) become significant since 2020 when considering windows of two seasons, being included in some of the final models with a positive coefficient. The differential in intercepts (INTERCEPTS\_L\-4\_CSUM\_DIFF) is consistently significant although not included in the final model between the 2021-23 seasons, with a positive coefficient.

Additional features that are significant when considering windows of two seasons include the differentials in metres gained (METRES\_GAINED\_L4\_CSUM\_DIFF), individually significant between 2016-19 with a positive coefficient; contested possessions (POSSESSIONS\_CONTESTED\_L4\_CSUM\_DIFF), individually between 2018-22 with a positive coefficient; and tackles in the Forward 50 area (TACKLES\_INSI\-DE50\_L4\_CSUM\_DIFF) between 2020-22 with a positive coefficient. Likewise, differentials in the score launches (SCORE\_LAUNCHES\_L4\_CSUM\_DIFF) are consistently significant at the individual level since 2018 and are included in the model for the seasons 2022-23 with a positive coefficient.

When modelled as season cumulatives, the differential in cumulative goal shots (GOALS\_SHOTS\_CSUM\_DIFF) is significant with a positive coefficient in almost every season from 2018 to 2022, for windows of both one and two seasons, although it is included in the model trained in the 2017 season with a negative coefficient. Differentials in the indicators of actions in the Forward 50 often appear significant individually and in models for both windows, including entries (INSIDE50\_CSUM\_DIFF) and marks (MARKS\_INSIDE50\_CSUM\_DIFF) with a positive coefficient. Differentials in rebounds (REBOUND\_INSIDE50S\_CSUM\_DIFF) are significant for some seasons with a negative coefficient for both windows. Differentials in the score launches (SCORE\_LAUNCHES\_CSUM\_DIFF) are often significant and included in models with a positive coefficient.

When modelled as season cumulatives, the differential in cumulative goal shots (GOALS\_SHOTS\_CSUM\_DIFF) is significant with a positive coefficient in almost every season from 2018 to 2022, for windows of both one and two seasons, although it is included in the model trained in the 2017 season with a negative coefficient. Differentials in the indicators of actions in the Forward 50 often appear significant individually and in models for both windows, including entries (INSIDE50\_CSUM\_DIFF) and marks (MARKS\_INSIDE50\_CSUM\_DIFF) with a positive coefficient. Differentials in rebounds (REBOUND\_INSIDE50S\_CSUM\_DIFF) are significant for some seasons with a negative coefficient for both windows. Differentials in the score launches (SCORE\_LAUNCHES\_CSUM\_DIFF) are often significant and included in models with a positive coefficient.

When considering windows of two seasons, the differentials in the lost opportunities to get the ball (CONTEST\_DEFENSIVE\_LOSS\_CSUM\_DIFF) appear to be consistently significant at the individual level with a negative coefficient between 2018-21. Likewise, differentials in the contested possessions at the ground level of the Forward 50 (GETS\_GROUNDBALL50\_CSUM\_DIFF) are individually significant, with a positive coefficient, for almost all windows between 2016-22. Finally, both the differentials in the metres gained (METRES\_GAINED\_CSUM\_DIFF) and tackles (TACKLES\_CSUM\_DIFF, TACKLES\_INSIDE50\_CSUM\_DIFF) are significant and with a positive coefficient, although the former indicator does not appear to be significant beyond the 2018-19 window.

\begin{table}[htbp]
    \centering
    \caption{Estimated coefficients for the time-variant Bradley-Terry fitted to windows of one season and with PIs encoded as cumulatives over the previous four games (2015-2023). Significant coefficients are in bold and underlined. Significant coefficients at the individual level only are in italics.
}
    \label{tab:bt_expansion_coefficients}
    \scalebox{0.53}{
    \begin{tabular}{lcccccccccc}
    \toprule
   &  \multicolumn{10}{c}{\textbf{Season}} \\
       \cmidrule{2-11}
     \textbf{FEATURE} & \textbf{2015} & \textbf{2016} & \textbf{2017} & \textbf{2018} & \textbf{2019} & \textbf{2020} & \textbf{2021} & \textbf{2022} & \textbf{2023}  & \textbf{2015-23}\\
    \midrule
   & \multicolumn{10}{c}{MATCH DIFFICULTY}\\
    \midrule 
    AT\_HOME & & \textbf{\underline{0.609}} & \textbf{\underline{0.404}} & & \textit{0.303} & & & \textit{0.521} & \textbf{\underline{0.527}} & \textbf{\underline{0.187}} \\
  HOMEGROUND &  \textit{0.381} & \textit{0.438} & & & \textbf{\underline{0.615}} & & & \textbf{\underline{0.716}} & \textit{0.409} & \textit{0.363} \\ INTERSTATE & \textbf{\underline{-0.545}} & \textit{-0.443} & \textit{-0.469} & & \textit{-0.530} & \textbf{\underline{-0.705}} & & \textit{-0.515} & \textit{-0.437} & \textbf{\underline{-0.268}} \\
      \midrule
   & \multicolumn{10}{c}{FORM}\\
    \midrule 
CONSECUTIVE\_LOSSES  &  &&&&&&\textbf{\underline{0.416}} && \textit{-0.164} & \textit{-0.139}\\ 
CONSECUTIVE\_WINS&  &&&&&& &&  & \textit{0.106}\\
L4G\_WINS  &  &&&&&&\textit{-0.280} && \textit{0.230} & \textit{0.299}\\
LADDER\_POSITION\_DIFF & \textit{0.039}&\textit{0.061}&\textit{0.038}&\textit{0.030}&&\textit{0.029}&&\textit{0.050}&&\textit{0.044}\\
LADDERLY\_POSITION\_DIFF & \textbf{\underline{0.047}}&\textbf{\underline{0.058}}&\textbf{\underline{0.025}}&\textbf{\underline{0.045}}&\textbf{\underline{0.031}}&\textbf{\underline{0.040}}&\textit{0.036}&&\textit{0.045}&\textbf{\underline{0.026}}\\
LG\_WON &&&&&&&&& \textbf{\underline{0.557}} &\textit{0.278}\\
PERCENTAGE\_DIFF &&& \textit{0.009} & \textit{0.007} &&& \textbf{\underline{0.011}} & \textit{0.012} && \textit{0.009}\\
POINTSAGAINST\_DIFF & &&  \textbf{\underline{0.002}} & & & &  \textit{0.002} & \textbf{\underline{0.002}} & \textbf{\underline{0.001}} & \textbf{\underline{0.001}} \\
POINTSFOR\_DIFF & \textbf{\underline{0.001}} & & & \textbf{\underline{0.002}} & \textit{0.003} & \textbf{\underline{0.003}} & & \textbf{\underline{0.002}} & \textit{0.001} & \textit{0.001} \\
      \midrule
   & \multicolumn{10}{c}{PIs}\\
    \midrule 
BOUNCES\_L4\_CSUM\_DIFF &&&&&&&&&&\textit{0.004}\\
CLANGERS\_L4\_CSUM\_DIFF  \\
CLEARANCES\_CENTRE\_L4\_CSUM\_DIFF &&&&&&&&&&\textit{0.009} \\
CLEARANCES\_L4\_CSUM\_DIFF&&&&&&&&&&\textit{0.006} \\
CLEARANCES\_STOPPAGE\_L4\_CSUM\_DIFF &&&&&&&\textit{0.012}&&&\textit{0.005}\\
CONTEST\_DEFENSIVE\_LOSS\_L4\_CSUM\_DIFF &&&&&\textit{-0.018}\\
CONTEST\_DEFENSIVE\_LOSS\_RATE\_L4\_CSUM\_DIFF&&&&& \textbf{\underline{-0.023}} &&&&& \textit{-0.008} \\
CONTEST\_OFFENSIVE\_WIN\_L4\_CSUM\_DIFF&&&&&&&&&&\textit{0.013}\\
CONTEST\_OFFENSIVE\_WIN\_RATE\_L4\_CSUM\_DIFF&&&&&&&&\textbf{\underline{0.025}}&& \\
DISPOSALS\_EFFECTIVE\_L4\_CSUM\_DIFF \\
DISPOSALS\_EFFICIENCY\_L4\_CSUM\_DIFF \\
DISPOSALS\_L4\_CSUM\_DIFF &&&&&&&&&&\textit{0.001}\\
FREES\_AGAINST\_L4\_CSUM\_DIFF \\
GETS\_GROUNDBALL\_L4\_CSUM\_DIFF &&&&&&&\textit{0.006}&&&\textit{0.002} \\
GETS\_GROUNDBALL50\_L4\_CSUM\_DIFF &&&&&&&\textit{0.018}&\textit{0.014}&&\textit{0.012}\\
GOALS\_ACCURACY\_L4\_CSUM\_DIFF \\
GOALS\_SHOTS\_L4\_CSUM\_DIFF &&&&\textit{0.020}&\textbf{\underline{0.016}}&\textit{0.015}&&&&\textit{0.013}\\
HANDBALLS\_L4\_CSUM\_DIFF \\
HITOUTS\_ADVANTAGE\_L4\_CSUM\_DIFF &&&&&&&&&\textbf{\underline{-0.013}}&\\
HITOUTS\_ADVANTAGE\_RATE\_L4\_CSUM\_DIFF&&&&&&&&&&\textit{0.005} \\
HITOUTS\_WIN\_RATE\_L4\_CSUM\_DIFF&\textbf{\underline{0.005}}&&&&&&&&\textit{-0.006}&\textit{0.002}  \\
INSIDE50\_L4\_CSUM\_DIFF&&\textbf{\underline{0.013}}&&\textit{0.014}&&&&&\textbf{\underline{0.009}}&\textbf{\underline{0.004}} \\
INTERCEPTS\_L4\_CSUM\_DIFF  &&&&&&&&&&\textit{0.005}\\
KICK2HANDBALL\_L4\_CSUM\_DIFF  &&&&&&&&&&\textit{0.400}\\
KICKS\_EFFECTIVE\_L4\_CSUM\_DIFF  &&&&&&&&&&\textit{0.001}\\
KICKS\_EFFICIENCY\_L4\_CSUM\_DIFF\\
KICKS\_L4\_CSUM\_DIFF &&&&\textit{0.004}&&&&&&\textit{0.002}\\
MARKS\_CONTESTED\_L4\_CSUM\_DIFF&&&&\textit{0.017}&&&&&&\textit{0.010}\\
MARKS\_INSIDE50\_L4\_CSUM\_DIFF&&&&&&&&&&\textit{0.010}\\
MARKS\_INTERCEPT\_L4\_CSUM\_DIFF&&&&&&&&&&\textit{0.009}\\
MARKS\_L4\_CSUM\_DIFF\\
MARKS\_ONLEAD\_L4\_CSUM\_DIFF\\
METRES\_GAINED\_L4\_CSUM\_DIFF&&&&\textit{0.000}&&&&&&\textit{0.000}\\
ONE\_PERCENTERS\_L4\_CSUM\_DIFF&&&&&&&&&&\textit{0.002}\\
POSSESSIONS\_CONTESTED\_L4\_CSUM\_DIFF&&&&&&&&\textbf{\underline{0.008}}&&\textit{0.004}\\
POSSESSIONS\_CONTESTED\_RATE\_L4\_CSUM\_DIFF&&&&&&&&&&\textit{0.019}\\
POSSESSIONS\_L4\_CSUM\_DIFF&&&&&&&&&&\textit{0.001}\\
POSSESSIONS\_UNCONTESTED\_L4\_CSUM\_DIFF\\
PRESSURE\_DEFENSEHALF\_L4\_CSUM\_DIFF& & \textit{-0.002} & & \textit{-0.003} &&&&&& \textit{-0.002}\\
PRESSURE\_L4\_CSUM\_DIFF\\
REBOUND\_INSIDE50S\_L4\_CSUM\_DIFF&\textit{-0.009}&&&&&&&&&\textit{-0.006}\\
SCORE\_LAUNCHES\_L4\_CSUM\_DIFF&&&&\textit{0.016}&\textit{0.012}&&&&\textit{0.012}&\textit{0.014}\\
SPOILS\_L4\_CSUM\_DIFF\\
TACKLES\_INSIDE50\_L4\_CSUM\_DIFF&&&&&&&&&&\textit{0.009}\\
TACKLES\_L4\_CSUM\_DIFF\\
TURNOVERS\_L4\_CSUM\_DIFF\\
\bottomrule
    \end{tabular}}
    \end{table}

Table \ref{tab:accuracy_variant} summarises the classification accuracies for the trained models, tested on all games of the following season. Compared to the models in the previous experiment, it seems that the fit is better, at least considering all available data, with the classification accuracy over the trained seasons reaching 68.54\% for both encodings (Table \ref{tab:accuracy_variantw2}). The fit for individual seasons or windows is often worse than that of the models in Experiment 2. At the season level, both encodings achieve similar accuracies over the training data in both windows.  As to the classification accuracy over the test seasons, the encoding for the last four games is only better until 2019 for windows of one season, where it is surpassed by the models with the season cumulatives, reaching accuracies as high as 69.38\% for 2020. For windows of two seasons, both encodings achieve similar accuracies over the test data, although some of these models improve the accuracy of those trained on single seasons.

\begin{table}
    \centering
    \caption{Classification accuracy for the Bradley-Terry team-specific, time-variant expansion, fitted to windows of one season (2015-2023) with both encodings of PIs.}
    \label{tab:accuracy_variant}
        \scalebox{0.6}{
    \begin{tabular}{c c c c cc}
    \toprule
    \multicolumn{6}{c}{Window: 1 season}\\
    \midrule
    && \multicolumn{2}{c}{Last 4 games cumulative } &\multicolumn{2}{c}{Season cumulative}\\
    \cmidrule(lr){3-4} \cmidrule(lr){5-6}
 Train Season & Test Season & Train Accuracy & Test Accuracy & Train Accuracy & Test Accuracy \\
 \cmidrule(lr){1-2}    \cmidrule(lr){3-4} \cmidrule(lr){5-6}
    2015 & 2016 & 71.08\% & 67.15\% & 67.16\% & 62.80\% \\
    2016 & 2017 & 71.50\% & 60.29\% & 74.40\% & 59.31\% \\
    2017 & 2018 &  62.75\% & 66.02\% & 70.10\% & 46.60\% \\
    2018 & 2019 & 71.36\% & 61.84\% & 65.53\% & 59.00\% \\
    2019 & 2020 & 70.05\% & 65.62\% & 66.18\% & 69.38\% \\
    2020 & 2021 & 68.75\% & 62.25\% & 73.12\% & 65.20\%\\
    2021 & 2022 & 68.63\% & 61.17\% & 66.67\% & 58.25\% \\
    2022 & 2023 & 72.82\% & 64.02\% & 72.33\% & 64.95\% \\
    2023 & & 70.56 & & 69.16\% \\
    \bottomrule
    \end{tabular}}
\end{table}

\subsection{Round-by-Round Prediction Accuracy (Experiment 4)}

Table \ref{tab:accuracy_variant_one} compares the classification accuracy of the strategies mentioned Section \ref{sec:strategy} with the results obtained from the already discussed models. When encoding PIs as cumulatives of the last four games, the Majority Voting strategy achieves the highest classification accuracy in seasons 2017, 2021 and 2022, and a high of 69.57\% for 2016. The Addition strategy achieves the best accuracies in seasons 2018, 2019 and 2019, while the highest accuracy (70.05\%) is achieved by the Incremental model in season 2015, although it performs considerably worse for the remaining seasons. The Substitution and Contest-Specific models never achieve the highest accuracies.

\begin{table}
    \centering
    \caption{Classification accuracy for the Bradley-Terry team-specific, time-variant (TS-TV) expansion, fitted to windows of one season (2015-2023) with both encodings of PIs. The highest accuracies for each season are underlined and in bold.}
    \label{tab:accuracy_variant_one}
        \scalebox{0.6}{
    \begin{tabular}{c c c c cccc}
    \toprule
    && &\multicolumn{5}{c}{Last 4 games cumulative }\\
    \cmidrule(lr){4-8} 
    Train Season & Test Season & Contest-Specific & TS-TV & Addition & Substitution & Incremental & Majority Voting\\
   \cmidrule(lr){1-3} \cmidrule(lr){4-8}
    2015 & 2016 & 63.93\% & 67.15\% & 66.18\% & 67.15\% & \textbf{\underline{70.05\%}}& 69.57\%\\
    2016 & 2017 & 62.43\% & 60.29\% & 61.76\% & 59.31\% & 51.96\% & \textbf{\underline{63.76\%}}\\
    2017 & 2018 &  62.50\% & 66.02\% & \textbf{\underline{67.48\%}} & 61.65\% & 64.56\% & 64.56\% \\
    2018 & 2019 & 59.46\% & 61.84\% &\textbf{\underline{66.18\%}}&62.80\%& 58.94\% &65.70\%\\
    2019 & 2020 & 63.12\% & \textbf{\underline{65.62\%}} & 65.00\% & 60.00\% & 55.63\% & 61.88\% \\
    2020 & 2021 & 60.77\% & 62.25\% & 61.27\% & 60.78\% & 57.85\% & \textbf{\underline{63.73\%}}\\
    2021 & 2022 & 58.70\% & 61.17\% & 61.65\% & 60.68\% & 60.68\% & \textbf{\underline{65.05\%}}\\
    2022 & 2023 & 67.37\% & 64.02\% & \textbf{\underline{67.76\%}} & 64.49\% & 53.27\% & 64.49\% \\
\midrule
&& &\multicolumn{5}{c}{Season cumulative }\\
    \cmidrule(lr){4-8} 
    Train Season & Test Season & Contest-Specific & TS-TV & Addition & Substitution & Incremental & Majority Voting\\
   \cmidrule(lr){1-3} \cmidrule(lr){4-8}
    2015 & 2016 & 63.93\% & 62.80\% & 67.15\% & 64.25\% & \textbf{\underline{71.50\%}}& 65.22\%\\
    2016 & 2017 & \textbf{\underline{62.43\%}} & 59.31\% & 57.35\% & 61.27\% & 50.49\% & 60.29\%\\
    2017 & 2018 &  62.50\% & 46.60\% & 65.53\% & 58.74\% & \textbf{\underline{66.50\%}} & 64.08\% \\
    2018 & 2019 & 59.46\% & 59.90\% &\textbf{\underline{64.25\%}}&62.32\%& 57.49\% &\textbf{\underline{64.25\%}}\\
    2019 & 2020 & 63.12\% & \textbf{\underline{69.38\%}} & 68.12\% & 61.88\% & 60.63\% & 66.25\% \\
    2020 & 2021 & 60.77\% & \textbf{\underline{65.20\%}} & 60.29\% & 55.39\% & 50.49\% & 60.78\%\\
    2021 & 2022 & 58.70\% & 58.25\% & 66.99\% & 65.06\% & 63.11\% & \textbf{\underline{67.96\%}}\\
    2022 & 2023 & \textbf{\underline{67.37\%}} & 64.95\% & 62.62\% & 61.21\% & 60.75\% & 62.62\% \\
    \bottomrule
    \end{tabular}}
\end{table}

In the season cumulative encoding, the Majority Voting model achieves worse results, although it performs the best for 2019 and 2022. Models predicting the full season perform best in four seasons, with the model from Experiment 3 performing better than any model from the previous encoding in seasons 2020 and 2021. The Incremental strategy achieves the highest accuracy for season 2015 (71.5\%) while also performing the worst in season 2021 with 50.49\%. 

The number of correctly predicted matches in the Finals Series was also computed and summarised in Table \ref{tab:finalsseries_variant_one}. With some exceptions for seasons 2019 and 2022, it seems that the models struggle to predict these games compared to those in the Home-and-Away season.

As to the teams, the models can correctly predict most of the outcomes for some of the weakest and strongest teams. For instance, predictions for North Melbourne are particularly more accurate than for other teams between 2021-23, a period where the team only won nine games in total. This is also the case for the Brisbane Lions (2015-18) and Richmond (2018-19). However, there are cases where the predictions for the ‘average’ teams are more accurate, meaning that, at this stage, the ability of these expansions to predict some teams better than others is inconclusive (Table \ref{tab:accuracyteams}).

\section{Discussion}
\label{sec:discussion}
\subsection{Summary and discussion of findings}


The findings for the standard BT model confirm that, while game outcomes alone are not enough to explain why a team is significantly stronger than another, the BT model can successfully identify teams that are significantly stronger or weaker than the average. For instance, it is able to capture North Melbourne’s poor performance between 2021-23, the Brisbane Lions’ evolution from poor to better results, and the best and worst teams in the 2023 season. Increasing the windows to cover more seasons captures performances that are significantly better in the medium term, like Richmond’s between seasons 2017-20, although windows with a higher number of seasons risk averaging out evolving performances, like those of the Brisbane Lions or West Coast. This implies that team strengths change gradually in the medium term, possibly in a window of three seasons, and that advantages for weaker teams in the draft are not necessarily reflected in the subsequent seasons.

The worsening predictive ability of the standard BT model as the windows enlarge is natural given that the model only considers outcomes, thereby missing cross-seasonal changes in teams’ performance which might be continued in the predicted season.

The significance of the AT\_HOME effect echoes the findings in \citet{lazarus2018factors} and \citet{robertson2018evaluating}, meaning playing as the Home Team creates a positive advantage for the team. Since the inclusion of the effect results in slight changes to the strength estimates, it is safe to assume that Home teams enjoy the advantage regardless of their strength. The increase in accuracy for both the train and test seasons confirms that the AT\_HOME advantage is explanatory and predictive of game outcomes.

The significance of the other difficulty features in Experiment 3 regardless of the encoding of other features suggests that they are independent from other effects. In particular, playing at their home venue creates an advantage for the team, while playing interstate results in a significant disadvantage, supporting the findings in \citet{robertson2018evaluating}.

In terms of form, while the difference in the ladder position is estimated to grant an advantage to teams placing higher on the ladder at the time of the game, the effect of the differential in the last ladder position of the previous season seems to grant a stronger advantage, coinciding with the findings of \citet{robertson2018evaluating} and \citet{fahey2019multifactorial}. This was expected after Experiments 1 and 2 confirmed that strengths change gradually, meaning the previous season’s performance will still be somehow predictive of the current season. Having the advantage on the points in favour and against are both associated with higher chances of winning, indicating a team’s offensive and defensive strength respectively.

The significance of the majority of the performance indicators regardless of their encoding in the model trained on all available data hints that the cumulative differentials in performance indicators are significantly different for winning and losing teams, even if they are not significant in multivariate models. The findings contribute to the literature on the predictive ability of these indicators encoded as differentials for forward game prediction.

In particular, having an advantage over the activity in the Forward 50 area, both during the season and the last four games, is indicative of a team having greater chances of winning. Indeed, a team with more entries, possessions, marks and tackles in the Forward 50 can be considered stronger as it is more likely to get close to the opponent's goal posts, get the football or take a mark (and its consequent free kick) and turn that into a goal or a behind. Moreover, teams that have an advantage in the shots at goal and scoring chains are naturally more likely to win since they are more likely than the other team to score. Other significant performance indicators were the metres gained, hinting that teams where players move more are more likely to win, perhaps because it is indicative of them moving across the field into the Forward 50; and the rebounds, with weaker teams having higher rebounds possibly because they are more likely to have moved the ball from their Defensive area to the Midfield, indicating they were pressured by the opposing team.

\subsection{Changing coefficients and eras}

Something noticeable about the results in Experiment 3 is that the effect of some performance indicators seems to change direction or are counterintuitive. For instance, having the advantage over the cumulative wins was associated with a decrease in the winning chance for seasons 2017, 2018 and the 2019-20 and 2022-23 windows. This can be partly explained by the modelling strategy and the fixtures in the AFL.

It is possible that the early stages of the seasons, where the team differentials are lower, are skewing the results. Moreover, if a strong team by the end of the season loses a game, their advantage in their differentials will also distort the direction of the performance indicator. Indeed, it may well be possible that a losing team has the advantage in cumulative wins if it faced weaker teams in the earlier rounds, or their opponent faced stronger teams.

Moreover, the inclusion of the Finals Series in the modelling, where only the strongest teams face each other might also contribute to the changing coefficients, especially if the teams that placed lower on the ladder manage to make it far into the Series. 

The findings also contribute to the discussion of ‘eras’ presented in the literature review. In particular, it seems that teams with an advantage in scoring (points in favour, goal shots, score launches) were more likely to win their subsequent matches between seasons 2018-20. This implies that weaker teams in this era are more likely to win over stronger teams if they have an advantage in scoring differentials.

Furthermore, being able to pressure the opponent in the Forward 50 seems to gain relevance from season 2020 onwards, with stronger teams having the advantage in the entries and groundballs, along with higher season cumulative in marks and tackles in this area. Both ‘eras’ potentially identify teams prioritising offensive strategies, with the first one focusing on scoring and the second one favouring a stronger presence in the Forward 50.

\subsection{Predictive ability of the Bradley-Terry expansions}

When considering all available data, the fit of the Bradley-Terry model significantly improves with the inclusion of team-specific, time-variant features known before the game. Their predictive ability also improves compared to the standard specification and the expansion allowing for contest-specific effects, hinting that, in the AFL, outcomes can be predicted by a team’s strength as explained by its form, cumulative performance indicators and match difficulty.

Although dependent on the season and the prediction strategy, the Bradley-Terry expansions can successfully predict up to 71.5\% of the games. Predicting games round-by-round, allowing for models to be updated or trained iteratively on the predicted season, often results in higher accuracies over the predicted games, although the models predicting the full season, whether it is with contest-specific effects or team-specific, time-variant features, still perform the best on some occasions. Strategies combining the previous models by majority voting result in stable accuracies above 60\%, reaching up to 67.96\%.

In conclusion, it appears that interpretable models like BT models might be feasible to predict game results in advance and design data-driven strategies in the AFL.

\subsection{Limitations and future lines of research}

The results are limited to the male competition of the AFL, seasons 2015 to 2023, and it could be possible that the features found to be significant do not correspond to those in earlier seasons, the female competition (AFLW) or national leagues of football.

Moreover, the worsening accuracy in the Finals Series might imply that the inclusion of these games for modelling is not optimal: the Series pitches the strongest teams, so it might not be representative of an average game. However, given that the Finals only account for nine games, it is complicated to model them separately without experiencing the disadvantages of a reduced sample size.

Given the better predictive performance on significantly stronger and weaker teams, it would be interesting to confirm whether the BT model with team-specific, time-variant features struggles to predict games pitching teams that are not significantly stronger or weaker. So far, results are inconclusive because they might not be related to the overall strength of the team but due to inter-seasonal changes in strengths (e.g.: teams performing significantly different than in the season before) or even winning and losing streaks. Clarifying this aspect would help review the modelling approaches or encoding of variables so that the models can predict teams equally well regardless of their strength.

Regarding the use of the BT models, it would be rational to study their performance on the backward prediction of games, this is, predicting game outcomes after the match takes place, so as to compare the model with the existing ones in the literature. Moreover, the results of the fitted models are promising enough to study the performance of the BT expansions with other combinations of predictions beyond majority voting, or new features from commercial providers, like network data. Finally, it would also be interesting to replicate the analysis in this thesis with other machine learning models to confirm if the same features are found to be significant and compare their predictive performance. From a more technical point of view, it would also be interesting to investigate the use of more general BT models, for instance those that embed model selection using LASSO regularization \citep[for instance, using the \texttt{BTLLasso} R package of][]{schauberger2019btllasso}, or mixed-effect models.

On another note, the BT expansions are limited to modelling game outcomes. AF, however, differs from other sports in the scoring, as teams can reach well beyond 100 points in a game, with score margins also ranging from a few points to dozens. BT models simply ignore this granularity in the results, so it would be interesting to model the score margin or the number of goals and behinds with pre-game variables. Hierarchical Bayesian \citep{manderson2018dynamic} and Poisson \citep{tsokos2019modeling} models have already been applied to predict scoring and margins in sports, so it would be worth examining whether these features can be predicted before the game and whether they improve our results.

\section{Conclusions}
\label{sec:conclusions}
Our findings showcase that the various BT models can be used to predict games' outcomes in the AFL using only pre-game data. Furthermore, they provide an interpretable understanding of team strengths and the impact of different pre-game features on match outcomes.

The model can identify the teams that are significantly stronger or weaker, both in single years and windows of multiple seasons. The significance of strengths across these windows, along with the advantage generated by a higher ladder position in the season before, suggests that team performance does not change drastically year-to-year, but gradually in a window of about three years.

Match difficulty features are confirmed to affect the winning chances of a team independently of its strength, with playing as the Home Team and in the home ground increasing them and playing interstate decreasing them. Teams with the advantage over the points in favour and against are also considered stronger. Finally, the differentials in cumulatives of performance indicators can be indicative of winning teams, in particular, leading the activity in the Forward 50, goal shots and scoring.


\bibliographystyle{elsarticle-harv} 
\bibliography{bib}

\appendix

\section{Summary of works on machine learning approaches in AF}

\label{appendixa}
\begin{table}[H]
    \centering
    \label{lit1}
    \scalebox{0.53}{
\begin{tabular}{p{5cm}p{3cm}p{4.9cm}p{4.9cm}p{8.2cm}}
    \toprule
\textbf{Work} & \textbf{Models} & \textbf{Features} & \textbf{Data} & \textbf{Comments}\\
\midrule 
\citet{braham2018complex} &
Linear Regression &
Differentials in mean betweenness last 4 games, outdegree &
2014 AFL Season (207 games). From Champion Data & 
Games of AF can be modelled as passing networks, with players as nodes and weighted edges as passes \\
\citet{fahey2019multifactorial} &
GLM with elastic net regularization &
 Differentials ladder position last season, player-based and team-based rating. Pre-game features &
2013-18 AFL Season (1,241 games). From fitzRoy package.
Player data from Champion Data&
Classify features as team-related and game-related.
Non-linear models and approaches including feature selection perform better\\
\citet{fransen2022cooperative} &
Binomial generalized linear mixed effects regression&
Linear combinations of network variables representing connectedness, in-degree variability, out-degree variability&
Network data from Champion Data (1,629 observations) &
Combine network measures of passing networks as connectedness, in-degree variability and out-degree variability. One-unit increases in connectedness increase the probability of winning by 5.3\%  \\
\citet{lane2020characterisation} & Calculate z-scores of offensive and defensive variables&Offensive (inside 50s, metres gained, etc.) and defensive features (tackles, rebound 50s, clangers, etc.)& 
Performance Indicator Data from AFL seasons 1999-2019 &
Identify eras in the AFL with significant differences in PIs. They identify an era of an offensive game (2007-08) moving to a defensive game (until 2011) with more tackles and disposals but lower scores \\
\citet{lazarus2018factors} &
Logistic mixed-effect regression & Playing Home/Away, travelling, age and weight differentials, days between games.&
Data for AFL seasons 2000-13 (5,109 games). From AFL Tables & Older and heavier teams are found to have significantly yet slightly higher chances of winning.
Playing as the away team significantly yet slightly decreases the chances of winning \\
\citet{manderson2018dynamic} &
Bayesian hierarchical model with Skellam distribution &
Score, goals and behinds for each team.&
Home-and-Away data from AFL seasons 2013-15. From Footywire &
Pre-game prediction of the number of goals and behinds separately.  It performed slightly worse in the last games of the season \\
\citet{robertson2016explaining} &
Logistic regression &Differentials in Performance Indicators (kicks, goal conversions, inside 50s, etc.)&
Performance Indicator data from AFL 2013-14 Season (396 games). Unspecified freely available sources & 81\% of teams with a kick differential above -1 won their games. 90\% of teams with higher additional kicks and goal conversion won their games.
Differentials in the inside 50s, marks, marks inside 50s and contested possessions were the most important predictors.\\
\citet{robertson2018evaluating} &
Logistic regression &
Number of wins over the past 4 games, and differentials in ladder rankings from the current and previous seasons &
AFL difficulty factors from the 2014 pre-season (198 games). From Champion Data and AFL Stats & Teams playing away show decreased chances of winning, especially if they play interstate\\
\citet{sargent2013evaluating} &
Linear regression &
Team ratings based on eigenvector centrality&
Games from AFL 2011 season, Geelong Football Club & Higher team connectedness increases the score margin.
The score margin can be predicted with pre-game connectedness values, to the point that the individual impact of players can be estimated \\
\citet{sargent2022sleep} &Linear mixed effects model&Sleep time, sleep efficiency, time between sleeping and waking up& Sleep data from 37 male AFL players &
Confirm that players sleep less when playing interstate. Mention the lack of studies relating poorer sleep to game outcomes \\
\citet{woods2017evolution} &Non-metric multidimensional scaling& Handballs, disposals, uncontested possessions, clangers, marks& Performance Indicators from AFL 2001-15 seasons. From Champion Data &
Identify eras in the AFL: a ‘possession’ era (2005-09 seasons), a ‘defensive’ era (2010-13) and a ‘repossession’ (2014-onwards) \\
\citet{young2019understanding} &
Parametric tests for network measures &
 Edge count, edge density, average path length, degree centrality, eigenvector centrality, betweenness centrality, transitivity&
Games from AFL 2009-16 seasons (1,516 games). From Champion Data&
Create passing networks based on player interactions. No significant differences are found in the betweenness centrality. Other network measures were significant but only slightly correlated to the score margin.\\
\citet{young2019modelling} &
Random Forest &
Differentials in the time in possession and meters gained, along with relative disposals, Inside 50s, Marks Inside 50, along the regular Marks and Contested Possessions &
Performance Indicator data from AFL 2001-16 seasons. From Champion Data &
Raise concerns about the lack of consensus on the identification of eras in the AFL\\
\citet{young2020understanding} &
Decision trees and GLMs &
Differentials in metres gained, time in possession, kicks &
Performance Indicator data (52 PIs)  from AFL 2009-16 seasons (1,516 games). From Champion Data. 14 Tactical Indicators form event data representing passes &
Tactical features are more important but they do not increase accuracy \\

\bottomrule
    \end{tabular}}
    \end{table}

\newpage
\section{Glossary of the used features}
\label{glossary}
\begin{table}[H]
    \centering
\caption{Glossary of the used features - Part I.}
    \label{tab2}
    \scalebox{0.53}{
\begin{tabular}{p{4.4cm}p{10.3cm}p{12.4cm}}
    \toprule
\textbf{Type}  & \textbf{Variable}    & \textbf{Description}                    \\
    \midrule
TARGET &
WIN&
Indicates whether the team won (1) or lost (0)\\
TARGET &
AWAY\_WIN&
Indicates whether the Away team won (1) or lost (0)\\
TARGET&
HOME\_WIN&
Indicates whether the Home team won (1) or lost (0)\\
MATCH DIFFICULTY &
AT\_HOME &
Indicates whether the team was designated as the Home Team (1) or the Away Team (0)\\
MATCH DIFFICULTY&
HOMEGROUND&
Indicates whether the team was played at their Home Ground (1) or not (0)\\
MATCH DIFFICULTY &
INTERSTATE &
Indicates whether the team was played outside their State (1) or not (0)\\
FORM &
CONSECUTIVE\_LOSSES &
Number of consecutive losses \\
FORM &
CONSECUTIVE\_WINS &
Number of consecutive wins\\
FORM &
L4G\_WINS&
Number of wins over the previous 4 games\\
FORM&
LADDER\_POSITION\_DIFF&
Differential in ladder position compared to the other team at the time of the game\\
FORM&
LADDERLY\_POSITION\_DIFF&
Differential in ladder position of last year compared to the other team\\
FORM&
LG\_WON&
Indicates whether the team won their previous game (1) or not (0)\\
FORM&
PERCENTAGE\_DIFF&
Differential in Points For to Points Against ratio compared to the other team\\
FORM&
POINTSAGAINST\_DIFF &
Differential in cumulative points scored against the team compared to the other team\\
FORM&
POINTSFOR\_DIFF &
Differential in cumulative points scored in favour of the team compared to the other team\\
FORM &
WINS\_CUMULATIVE\_DIFF&
Differential in cumulative wins compared to the other team\\
PI &
BOUNCES\_L4\_CSUM\_DIFF &
Differential in the cumulative sum of bounces over the previous 4 games, compared to the other team\\
PI &
CLANGERS\_L4\_CSUM\_DIFF &
Differential in the cumulative sum of clangers (mistakes) over the previous 4 games, compared to the other team\\
PI &
CLEARANCES\_CENTRE\_L4\_CSUM\_DIFF &
Differential in the cumulative sum of centre clearances (clearing the centre area) over the previous 4 games, compared to the other team\\
PI&
CLEARANCES\_L4\_CSUM\_DIFF&
Differential in the cumulative sum of clearances (clearing the centre or stoppage area) over the previous 4 games, compared to the other team \\
PI&
CLEARANCES\_STOPPAGE\_L4\_CSUM\_DIFF &
Differential in the cumulative sum of stoppage clearances (clearing the stoppage area) over the previous 4 games, compared to the other team \\
PI &
CONTEST\_DEFENSIVE\_LOSS\_L4\_CSUM\_DIFF &
Differential in the cumulative losses of one-to-one contests (moments where two opposing players can get the ball) over the previous 4 games, compared to the other team \\
PI &
CONTEST\_DEFENSIVE\_LOSS\_RATE\_L4\_CSUM\_DIFF &
Differential in the rate of losses of one-to-one contests (moments where two opposing players can get the ball) over the previous 4 games, compared to the other team \\
PI &
CONTEST\_OFFENSIVE\_WIN\_L4\_CSUM\_DIFF&
Differential in the cumulative wins of one-to-one contests (moments where two opposing players can get the ball) over the previous 4 games, compared to the other team\\
PI &
CONTEST\_OFFENSIVE\_WIN\_RATE\_L4\_CSUM\_DIFF &
Differential in the rate of wins of one-to-one contests (moments where two opposing players can get the ball) over the previous 4 games, compared to the other team \\
PI &
DISPOSALS\_EFFECTIVE\_L4\_CSUM\_DIFF &
Differential in the cumulative sum of successful disposals (kicks and handballs) over the previous 4 games, compared to the other team\\
PI&
DISPOSALS\_EFFICIENCY\_L4\_CSUM\_DIFF &
Differential in the rate of successful disposals (kicks and handballs) over the previous 4 games, compared to the other team\\
PI &
DISPOSALS\_L4\_CSUM\_DIFF &
Differential in the cumulative sum of disposals (kicks and handballs) over the previous 4 games, compared to the other team\\
PI &
FREES\_AGAINST\_L4\_CSUM\_DIFF &
Differential in the cumulative sum of free kicks against the team over the previous 4 games, compared to the other team\\
PI &
GETS\_GROUNDBALL\_L4\_CSUM\_DIFF &
Differential in the cumulative sum of contested possessions won in the ground over the previous 4 games, compared to the other team\\
PI &
GETS\_GROUNDBALL50\_L4\_CSUM\_DIFF &
Differential in the cumulative sum of contested possessions won in the ground of the Forward 50 area over the previous 4 games, compared to the other team\\
PI &
GOALS\_ACCURACY\_L4\_CSUM\_DIFF &
Differential in the goals-to-goal shots ratio over the previous 4 games, compared to the other team\\
\bottomrule
\end{tabular}}
\end{table}

\begin{table}[H]
    \centering
\caption{Glossary of the used features - Part II.}
    \label{tab3}
    \scalebox{0.53}{
\begin{tabular}{p{4.4cm}p{10.3cm}p{12.4cm}}
    \toprule
\textbf{Type}  & \textbf{Variable}    & \textbf{Description}                    \\
    \midrule
PI&
GOALS\_SHOTS\_L4\_CSUM\_DIFF&
Differential in the cumulative sum of goals over the previous 4 games, compared to the other team\\
PI&
HANDBALLS\_L4\_CSUM\_DIFF &
Differential in the cumulative sum of handballs over the previous 4 games, compared to the other team\\
PI &
HITOUTS\_ADVANTAGE\_L4\_CSUM\_DIFF &
Differential in the cumulative sum of hitouts to advantage (knocking the ball to a teammate in a contest after a stoppage) over the previous 4 games, compared to the other team \\
PI &
HITOUTS\_ADVANTAGE\_RATE\_L4\_CSUM\_DIFF &
Differential in the ratio of hitouts to advantage to hitouts over the previous 4 games, compared to the other team\\
PI &
HITOUTS\_WIN\_RATE\_L4\_CSUM\_DIFF &
Differential in the ratio of contests with a hitout win to contests over the previous 4 games, compared to the other team \\
PI &
INSIDE50\_L4\_CSUM\_DIFF &
Differential in the cumulative sum of entries into the Forward 50 area over the previous 4 games, compared to the other team \\
PI &
INTERCEPTS\_L4\_CSUM\_DIFF &
Differential in the cumulative sum of intercepts (taking the ball from the other team) over the previous 4 games, compared to the other team \\
PI &
KICK2HANDBALL\_L4\_CSUM\_DIFF &
Differential in the ratio of kicks to handballs over the previous 4 games, compared to the other team \\
PI &
KICKS\_EFFECTIVE\_L4\_CSUM\_DIFF &
Differential in the cumulative sum of successful kicks over the previous 4 games, compared to the other team \\
PI &
KICKS\_EFFICIENCY\_L4\_CSUM\_DIFF &
Differential in the ratio of effective kicks to all kicks over the previous 4 games, compared to the other team\\
PI &
KICKS\_L4\_CSUM\_DIFF &
Differential in the cumulative sum of kicks over the previous 4 games, compared to the other team \\
PI &
MARKS\_CONTESTED\_L4\_CSUM\_DIFF &
Differential in the cumulative sum of marks taken under pressure over the previous 4 games, compared to the other team \\
PI &
MARKS\_INSIDE50\_L4\_CSUM\_DIFF &
Differential in the cumulative sum of marks taken in the Forward 50 area over the previous 4 games, compared to the other team \\
PI &
MARKS\_INTERCEPT\_L4\_CSUM\_DIFF &
Differential in the cumulative sum of marks taken from the opponent over the previous 4 games, compared to the other team \\
PI &
MARKS\_L4\_CSUM\_DIFF &
Differential in the cumulative sum of marks over the previous 4 games, compared to the other team \\
PI &
MARKS\_ONLEAD\_L4\_CSUM\_DIFF &
Differential in the cumulative sum of marks taken under no pressure over the previous 4 games, compared to the other team \\
PI &
METRES\_GAINED\_L4\_CSUM\_DIFF &
Differential in the cumulative sum of metres gained over the previous 4 games, compared to the other team\\
PI &
ONE\_PERCENTERS\_L4\_CSUM\_DIFF &
Differential in the cumulative sum of one-percenters (defensive but rare actions) over the previous 4 games, compared to the other team \\
PI &
POSSESSIONS\_CONTESTED\_L4\_CSUM\_DIFF &
Differential in the cumulative sum of contested possessions (possessions gained under dispute) over the previous 4 games, compared to the other team \\
PI &
POSSESSIONS\_CONTESTED\_RATE\_L4\_CSUM\_DIFF &
Differential in the ratio of possessions won in a dispute to all disputes over the previous 4 games, compared to the other team \\
PI &
POSSESSIONS\_L4\_CSUM\_DIFF &
Differential in the cumulative sum of possessions over the previous 4 games, compared to the other team \\
PI &
POSSESSIONS\_UNCONTESTED\_L4\_CSUM\_DIFF &
Differential in the cumulative sum of uncontested possessions (possessions gained under no pressure) over the previous 4 games, compared to the other team\\
PI &
PRESSURE\_DEFENSEHALF\_L4\_CSUM\_DIFF &
Differential in the cumulative sum of pressure acts in the Defense Half of the ground over the previous 4 games, compared to the other team \\
PI &
PRESSURE\_L4\_CSUM\_DIFF &
Differential in the cumulative sum of pressure acts over the previous 4 games, compared to the other team \\
PI &
REBOUND\_INSIDE50S\_L4\_CSUM\_DIFF &
Differential in the cumulative sum of rebound 50s (moving the ball from the Defensive 50 to the midfield) over the previous 4 games, compared to the other team\\
PI &
SCORE\_LAUNCHES\_L4\_CSUM\_DIFF &
Differential in the cumulative sum of score launches (scoring chains after successful clearances, intercepts, free kicks, hitouts) over the previous 4 games, compared to the other team \\
PI &
SPOILS\_L4\_CSUM\_DIFF &
Differential in the cumulative sum of spoils (preventing opponents from taking a mark) over the previous 4 games, compared to the other team\\
PI &
TACKLES\_INSIDE50\_L4\_CSUM\_DIFF &
Differential in the cumulative sum of tackles in the Forward 50 area over the previous 4 games, compared to the other team\\
PI &
TACKLES\_L4\_CSUM\_DIFF &
Differential in the cumulative sum of tackles over the previous 4 games, compared to the other team \\
PI &
TURNOVERS\_L4\_CSUM\_DIFF &
Differential in the cumulative sum of turnovers (losses of possession) over the previous 4 games, compared to the other team\\
\bottomrule
\end{tabular}}
\end{table}

\begin{table}[H]
    \centering
\caption{Glossary of the used features - Part III.}
    \label{tab4}
    \scalebox{0.53}{
\begin{tabular}{p{4.4cm}p{10.3cm}p{12.4cm}}
    \toprule
\textbf{Type}  & \textbf{Variable}    & \textbf{Description}                    \\
    \midrule
PI &
BOUNCES\_CSUM\_DIFF &
Differential in the cumulative sum of bounces over the season, compared to the other team.\\
PI &
CLANGERS\_CSUM\_DIFF &
Differential in the cumulative sum of clangers (mistakes) over the season, compared to the other team \\
PI &
CLEARANCES\_CENTRE\_CSUM\_DIFF &
Differential in the cumulative sum of centre clearances (clearing the centre area) over the season, compared to the other team \\
PI &
CLEARANCES\_CSUM\_DIFF &
Differential in the cumulative sum of clearances (clearing the centre or stoppage area) over the season, compared to the other team\\
PI &
CLEARANCES\_STOPPAGE\_CSUM\_DIFF &
Differential in the cumulative sum of stoppage clearances (clearing the stoppage area) over the season, compared to the other team\\
PI &
CONTEST\_DEFENSIVE\_LOSS\_CSUM\_DIFF &
Differential in the cumulative losses of one-to-one contests (moments where two opposing players can get the ball) over the season, compared to the other team\\
PI &
CONTEST\_DEFENSIVE\_LOSS\_RATE\_CSUM\_DIFF &
Differential in the rate of losses of one-to-one contests (moments where two opposing players can get the ball) over the season, compared to the other team \\
PI &
CONTEST\_OFFENSIVE\_WIN\_CSUM\_DIFF &
Differential in the cumulative wins of one-to-one contests (moments where two opposing players can get the ball) over the season, compared to the other team\\
PI &
CONTEST\_OFFENSIVE\_WIN\_RATE\_CSUM\_DIFF &
Differential in the rate of wins of one-to-one contests (moments where two opposing players can get the ball) over the season, compared to the other team\\
PI &
DISPOSALS\_CSUM\_DIFF &
Differential in the cumulative sum of successful disposals (kicks and handballs) over the season, compared to the other team \\
PI &
DISPOSALS\_EFFECTIVE\_CSUM\_DIFF &
Differential in the rate of successful disposals (kicks and handballs) over the season, compared to the other team\\
PI &
DISPOSALS\_EFFICIENCY\_CSUM\_DIFF &
Differential in the cumulative sum of disposals (kicks and handballs) over the season, compared to the other team\\
PI &
FREES\_AGAINST\_CSUM\_DIFF &
Differential in the cumulative sum of free kicks against the team over the season, compared to the other team \\
PI &
GETS\_GROUNDBALL\_CSUM\_DIFF &
Differential in the cumulative sum of contested possessions won in the ground over the season, compared to the other team\\
PI &
GETS\_GROUNDBALL50\_CSUM\_DIFF &
Differential in the cumulative sum of contested possessions won in the ground of the Forward 50 area over the season, compared to the other team\\
PI &
GOALS\_ACCURACY\_CSUM\_DIFF &
Differential in the goals-to-goal shots ratio over the season, compared to the other team\\
PI &
GOALS\_SHOTS\_CSUM\_DIFF &
Differential in the cumulative sum of goals over the season, compared to the other team\\
PI &
HANDBALLS\_CSUM\_DIFF &
Differential in the cumulative sum of handballs over the season, compared to the other team\\
PI &
HITOUTS\_ADVANTAGE\_CSUM\_DIFF &
Differential in the cumulative sum of hitouts to advantage (knocking the ball to a teammate in a contest after a stoppage) over the season, compared to the other team\\
PI &
HITOUTS\_ADVANTAGE\_RATE\_CSUM\_DIFF &
Differential in the ratio of hitouts to advantage to hitouts over the season, compared to the other team\\
PI &
HITOUTS\_WIN\_RATE\_CSUM\_DIFF &
Differential in the ratio of contests with a hitout win to contests over the season, compared to the other team \\
PI &
INSIDE50\_CSUM\_DIFF &
Differential in the cumulative sum of entries into the Forward 50 area over the season, compared to the other team \\
PI &
INTERCEPTS\_CSUM\_DIFF &
Differential in the cumulative sum of intercepts (taking the ball from the other team) over the season, compared to the other team \\
PI &
KICK2HANDBALL\_CSUM\_DIFF &
Differential in the ratio of kicks to handballs over the season, compared to the other team \\
PI &
KICKS\_CSUM\_DIFF &
Differential in the cumulative sum of kicks over the season, compared to the other team\\
PI &
KICKS\_EFFECTIVE\_CSUM\_DIFF &
Differential in the cumulative sum of successful kicks over the season, compared to the other team \\
PI &
KICKS\_EFFICIENCY\_CSUM\_DIFF &
Differential in the ratio of effective kicks to all kicks over the season, compared to the other team \\
PI &
MARKS\_CONTESTED\_CSUM\_DIFF &
Differential in the cumulative sum of marks taken under pressure over the season, compared to the other team\\
PI &
MARKS\_CSUM\_DIFF &
Differential in the cumulative sum of marks over the season, compared to the other team \\
\bottomrule
\end{tabular}}
\end{table}

\begin{table}[H]
    \centering
\caption{Glossary of the used features - Part IV.}
    \label{tab5}
    \scalebox{0.53}{
\begin{tabular}{p{4.4cm}p{10.3cm}p{12.4cm}}
    \toprule
\textbf{Type}  & \textbf{Variable}    & \textbf{Description}                    \\
    \midrule
PI &
MARKS\_INSIDE50\_CSUM\_DIFF &
Differential in the cumulative sum of marks taken in the Forward 50 area over the season, compared to the other team \\
PI &
MARKS\_INTERCEPT\_CSUM\_DIFF &
Differential in the cumulative sum of marks taken from the opponent over the season, compared to the other team \\
PI &
MARKS\_ONLEAD\_CSUM\_DIFF &
Differential in the cumulative sum of marks taken under no pressure over the season, compared to the other team \\
PI &
METRES\_GAINED\_CSUM\_DIFF &
Differential in the cumulative sum of metres gained over the season, compared to the other team \\
PI &
ONE\_PERCENTERS\_CSUM\_DIFF &
Differential in the cumulative sum of one-percenters (defensive but rare actions) over the season, compared to the other team \\
PI &
POSSESSIONS\_CONTESTED\_CSUM\_DIFF &
Differential in the cumulative sum of contested possessions (possessions gained under dispute) over the season, compared to the other team \\
PI &
POSSESSIONS\_CONTESTED\_RATE\_CSUM\_DIFF &
Differential in the ratio of possessions won in a dispute to all disputes over the season, compared to the other team\\
PI &
POSSESSIONS\_CSUM\_DIFF & 
Differential in the cumulative sum of possessions over the season, compared to the other team \\
PI &
POSSESSIONS\_UNCONTESTED\_CSUM\_DIFF &
Differential in the cumulative sum of uncontested possessions (possessions gained under no pressure) over the season, compared to the other team \\
PI &
PRESSURE\_CSUM\_DIFF &
Differential in the cumulative sum of pressure acts over the season, compared to the other team \\
PI &
PRESSURE\_DEFENSEHALF\_CSUM\_DIFF &
Differential in the cumulative sum of pressure acts in the Defense Half of the ground over the season, compared to the other team\\
PI &
REBOUND\_INSIDE50S\_CSUM\_DIFF &
Differential in the cumulative sum of rebound 50s (moving the ball from the Defensive 50 to the midfield) over the season, compared to the other team \\
PI &
SCORE\_LAUNCHES\_CSUM\_DIFF &
Differential in the cumulative sum of score launches (scoring chains after successful clearances, intercepts, free kicks, hitouts) over the season, compared to the other team \\
PI &
SPOILS\_CSUM\_DIFF &
Differential in the cumulative sum of spoils (preventing opponents from taking a mark) over the previous 4 games, compared to the other team \\
PI &
TACKLES\_CSUM\_DIFF &
Differential in the cumulative sum of tackles over the season, compared to the other team\\
PI &
TACKLES\_INSIDE50\_CSUM\_DIFF &
Differential in the cumulative sum of tackles in the Forward 50 area over the season, compared to the other team \\
PI &
TURNOVERS\_CSUM\_DIFF &
Differential in the cumulative sum of turnovers (losses of possession) over the season, compared to the other team \\
    \bottomrule
\end{tabular}}
\end{table}

\newpage 

\section{Data preparation}
\label{prep}
\begin{figure}[H]
    \centering
    \includegraphics[scale=0.6]{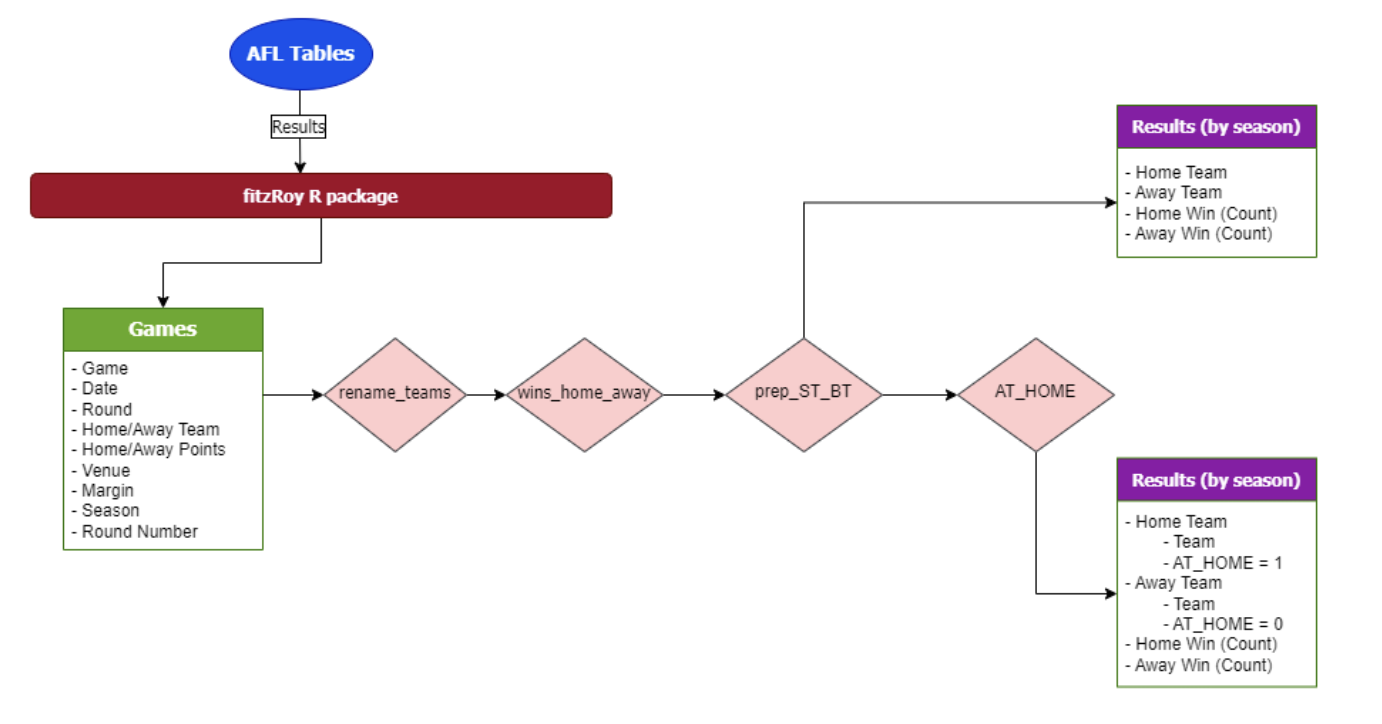}
    \caption{Data processing for standard and contest-specific Bradley-Terry models.}
    \label{fig:pipeline1}
\end{figure}

Following \citet{firth2012bradley}, game outcomes were turned into binomial frequencies, where each entry pairs the Home and the Away Team retrieving the absolute frequencies of wins for each team against the other. For instance, for the AFL data used in this thesis, there is an entry pairing the Brisbane Lions as the Home Team against Collingwood as the Away Team, and the number of wins each team achieved against the other with this pairing. Likewise, there will be another entry with Collingwood as the Home Team and the Brisbane Lions as the Away Team. Since the AFL pairs 18 teams through less than 30 rounds for our limited number of seasons, the binomial frequencies are expected to be low for all pairs. FIGURE  B.2.1. summarises the data processing for this expansion. The data is also expanded to include an AT\_HOME effect, which will be 1 for the Home Team and 0 for the Away Team, keeping the data design as binomial counts. This process is reflected in \ref{fig:pipeline1}. 

For models fitted using time-variant features,  the data has to be structured in the form of three data frames, since contests cannot be grouped by the Home and Away pairs: one with a set of teams playing Home, their features and results; another with the set of Away teams; and a third dataset with time-invariant features which, for this expansion, are none. All datasets were complete and contained no missing values.

\begin{figure}[H]
    \includegraphics[scale=0.5]{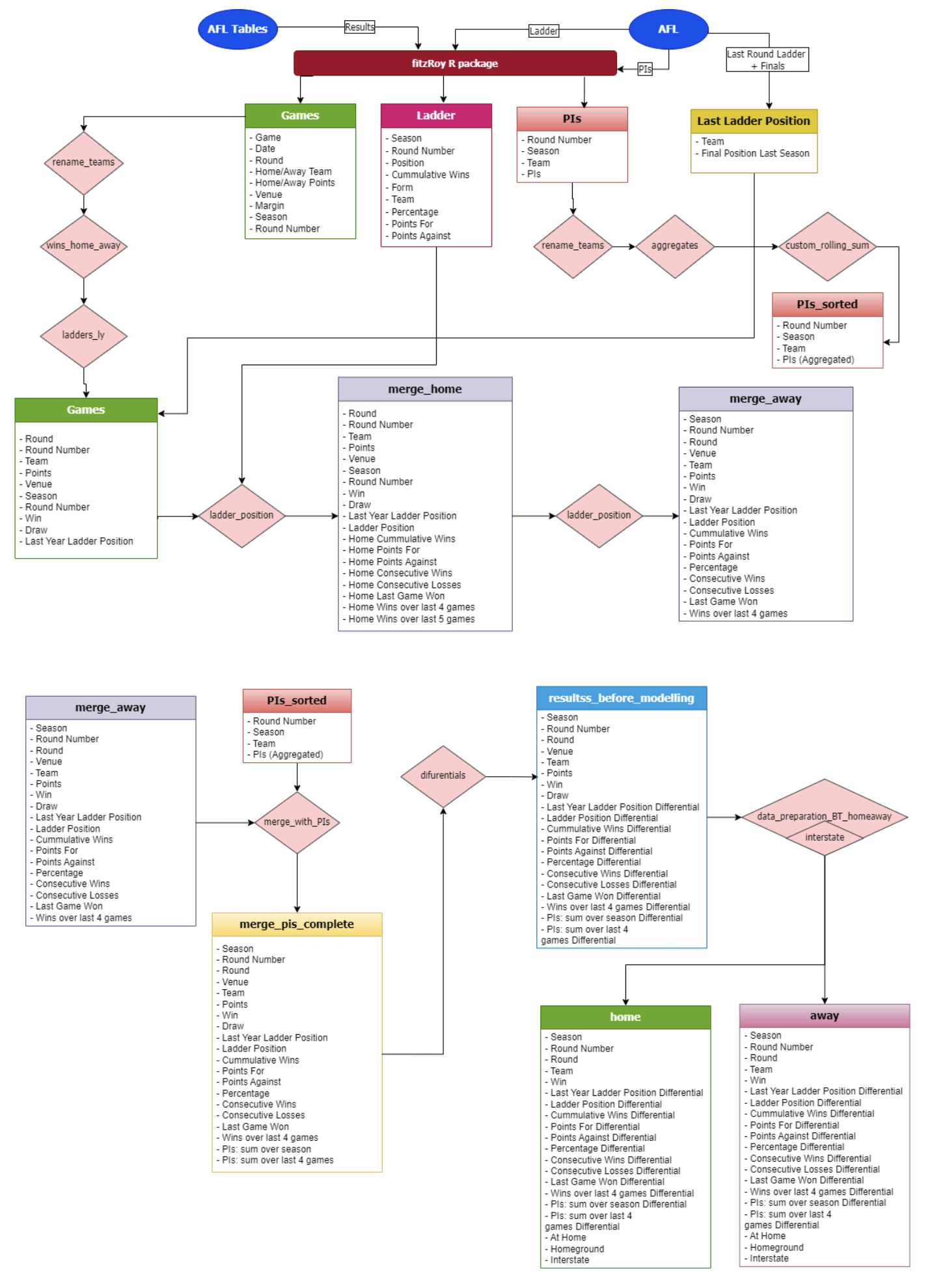}
    \caption{Data processing for the team-specific, time-variant Bradley-Terry model. Data from games, merge\_away and merge\_PIs\_complete contains entries from both Home and Away teams. Data from PIs, PIs\_sorted, ladder and last ladder position contains entries by round and team.}
    \label{fig:pipeline2}
\end{figure}

\section{Description of Experiments}

\newpage 

\section{Result Tables}

\begin{table} [H]
    \centering
    \caption{Estimated Strength Coefficients with the Standard Bradley-Terry Model, Fitted to Windows of Two Seasons (2015-2023). Coefficients found to be significant at the 5\% level are in bold and underlined. The coefficient for the reference team is 0.00.}
    \label{tab:coefficientsw2}
    \scalebox{0.7}{
    \begin{tabular}{l *{8}{c}}
    \toprule
    \multicolumn{1}{c}{\textbf{Team}} & \multicolumn{8}{c}{\textbf{Season}} \\
    \cmidrule(lr){2-9}
    & 2015-16 & 2016-17 & 2017-18 & 2018-19 & 2019-20 & 2020-21 & 2021-22 & 2022-23 \\
    \midrule
    Adelaide & 0.83 & \textbf{\underline{0.96}} & 0.42 & -0.03 & -0.58 & \textbf{\textbf{\underline{-1.13}}} & -0.83 & -0.50 \\
    Brisbane Lions & \textbf{\underline{-1.97}} & \textbf{\underline{-1.75}} & \textbf{\underline{-1.63}} & -0.20 & \textbf{\underline{1.20}} & \textbf{\underline{1.00}} & 0.81 & \textbf{\underline{0.94}} \\
    Carlton & \textbf{\underline{-1.40}} & \textbf{\underline{-1.11}} & \textbf{\underline{-1.92}} & \textbf{\underline{-1.58}} & -0.50 & -0.49 & -0.21 & 0.30  \\
    Collingwood & -0.40 & -0.45 & 0.02 & 0.84 & 0.72 & -0.36 & -0.02 & \textbf{\underline{1.14}} \\
    Essendon & \textbf{\underline{-1.61}} & \textbf{\underline{-0.93}} & -0.13 & 0.22 & 0.00 & -0.28 & -0.42 & -0.53 \\
    Fremantle & 0.00 & \textbf{\underline{-1.11}} & -0.84 & -0.50 & -0.21 & -0.26 & 0.34 & 0.21 \\
    Gold Coast & \textbf{\underline{-1.50}} & \textbf{\underline{-1.29}} & \textbf{\underline{-1.75}} & \textbf{\underline{-1.87}} & \textbf{\underline{-1.24}} & -0.81 & -0.57 & -0.53 \\
    Geelong & 0.84 & \textbf{\underline{0.97}} & 0.35 & 0.59 & \textbf{\underline{1.00}} & \textbf{\underline{0.96}} & \textbf{\underline{1.30}} & 0.67 \\
    Greater Western Sydney & 0.49 & 0.85 & 0.44 & 0.55 & 0.43 & 0.10 & -0.27 & -0.30 \\
    Hawthorn & \textbf{\underline{1.17}} & 0.44 & 0.00 & 0.32 & -0.20 & -0.81 & -0.70 & \textbf{\underline{-0.96}} \\
    Melbourne & -0.57 & -0.12 & 0.13 & -0.21 & -0.44 & \textbf{\underline{1.06}} & \textbf{\underline{1.33}} & 0.72 \\
    North Melbourne & 0.37 & -0.48 & -0.81 & -0.06 & -0.55 &  \textbf{\underline{-1.57}} & \textbf{\underline{-1.91}} & \textbf{\underline{-2.49}}\\
    Port Adelaide & 0.05 & 0.06 & 0.08 & 0.10 & 0.79 & \textbf{\underline{1.34}} & 0.54 & 0.29\\
    Richmond & -0.02 & 0.17 & \textbf{\underline{0.94}} & \textbf{\underline{1.34}} & \textbf{\underline{1.42}} & 0.55 & 0.06 & -0.04 \\
    Saint Kilda & -0.47 & 0.00 & -0.87 & \textbf{\underline{-0.99}} & 0.03 & 0.18 & 0.00 & 0.00\\
    Sydney & \textbf{\underline{1.04}} & 0.84 & 0.29 & 0.00 & -0.60 & -0.00 & 0.85 & 0.50 \\
    Western Bulldogs & 0.88 & 0.54 & -0.58 & -0.28 & 0.33 & 0.72 & 0.59 & 0.04 \\
    West Coast & \textbf{\underline{1.17}} & 0.46 & 0.45 & 1.00 & 0.87 & 0.21 & \textbf{\underline{-1.05}} & \textbf{\underline{-2.45}} \\
    \bottomrule
    \end{tabular}
}
\end{table}

\begin{table} [H]
    \centering
    \caption{Estimated Strength Coefficients with the Standard Bradley-Terry Model, Fitted to Windows of Three Seasons (2015-2023). Coefficients found to be significant at the 5\% level are in bold and underlined. The coefficient for the reference team is 0.00.}
    \label{tab:coefficientsw3}
    \scalebox{0.7}{
    \begin{tabular}{l *{7}{c}}
    \toprule
    \multicolumn{1}{c}{\textbf{Team}} & \multicolumn{7}{c}{\textbf{Season}} \\
    \cmidrule(lr){2-8}
    & 2015-17 & 2016-18 & 2017-19 & 2018-20 & 2019-21 & 2020-22 & 2021-23\\
    \midrule
    Adelaide & \textbf{\underline{0.94}} & 0.59 & 0.18 & -0.40 & -0.65 & \textbf{\textbf{\underline{-1.10}}} & -0.56 \\
    Brisbane Lions & \textbf{\underline{-1.58}} & \textbf{\underline{-1.73}} & \textbf{\underline{-0.70}} & 0.22 & \textbf{\underline{0.93}} & \textbf{\underline{0.84}} & \textbf{\underline{0.87}} \\
    Carlton & \textbf{\underline{-1.13}} & \textbf{\underline{-1.55}} & \textbf{\underline{-1.51}} & \textbf{\underline{-1.15}} & -0.52 & -0.37 & 0.07  \\
    Collingwood & -0.24 & -0.14 & 0.29 & 0.64 & 0.15 & -0.07 & 0.50 \\
    Essendon & \textbf{\underline{-0.82}} & -0.66 & -0.04 & 0.00 & -0.06 & -0.55 & -0.32 \\
    Fremantle  & -0.10 & \textbf{\underline{-1.06}} & -0.66 & -0.46 & -0.21 & 0.02 & 0.14 \\
    Gold Coast & \textbf{\underline{-1.20}} & \textbf{\underline{-1.53}} & \textbf{\underline{-1.72}} & \textbf{\underline{-1.49}} & \textbf{\underline{-1.05}} & -0.73 & -0.55 \\
    Geelong & \textbf{\underline{0.89}} & 0.61 & 0.48 & 0.62 & \textbf{\underline{0.89}} & \textbf{\underline{1.03}} & \textbf{\underline{0.79}} \\
    Greater Western Sydney  & 0.68 & 0.61 & 0.45 & 0.33 & 0.31 & -0.37 & -0.06 \\
    Hawthorn & \textbf{\underline{0.82}} & 0.35 & 0.00 & -0.02 & -0.38 & \textbf{\underline{-0.87}} & \textbf{\underline{-0.77}} \\
    Melbourne & -0.20 & 0.04 & -0.26 & -0.13  & 0.30 & \textbf{\underline{0.85}} & \textbf{\underline{1.04}} \\
    North Melbourne & 0.00 & -0.43 & -0.57 & -0.42 &  \textbf{\underline{-0.86}} & \textbf{\underline{-1.93}} & \textbf{\underline{-1.97}}\\
    Port Adelaide & 0.28 & 0.00 & 0.05 & 0.44 & \textbf{\underline{0.87}} & 0.66 & 0.63\\
    Richmond & 0.42 & 0.43 & \textbf{\underline{1.02}} & \textbf{\underline{1.24}} & \textbf{\underline{0.83}} & 0.34 & -0.01 \\
    Saint Kilda & -0.16 & -0.53 & \textbf{\underline{-0.76}} & -0.54 & 0.00 & 0.00 & 0.00\\
    Sydney & \textbf{\underline{0.94}} & 0.61 & 0.00 & -0.27 & -0.13 & 0.24 & 0.58 \\
    Western Bulldogs & 0.66 & 0.04 & -0.35 & -0.14 & 0.52 & 0.36 & 0.40 \\
    West Coast & \textbf{\underline{0.86}} & 0.58 & 0.53 & \textbf{\underline{0.88}} & 0.45 & -0.60 & \textbf{\underline{-1.36}} \\
    \bottomrule
    \end{tabular}
}
\end{table}

\begin{table} [H]
    \centering
    \caption{Estimated Strength Coefficients with the Standard Bradley-Terry Model, Fitted to Windows of Four Seasons (2015-2023). Coefficients found to be significant at the 5\% level are in bold and underlined. The coefficient for the reference team is 0.00.}
    \label{tab:coefficientsw4}
    \scalebox{0.7}{
    \begin{tabular}{l *{6}{c}}
    \toprule
    \multicolumn{1}{c}{\textbf{Team}} & \multicolumn{6}{c}{\textbf{Season}} \\
    \cmidrule(lr){2-7}
    & 2015-18 & 2016-19 & 2017-20 & 2018-21 & 2019-22 & 2020-23\\
    \midrule
    Adelaide & \textbf{\underline{0.73}} & 0.38 & 0.00 & -0.43 & \textbf{\textbf{\underline{-0.67}}} & \textbf{\underline{-0.75}} \\
    Brisbane Lions & \textbf{\underline{-1.54}} & \textbf{\underline{-0.93}} & -0.13 & 0.34 & \textbf{\underline{0.88}} & \textbf{\underline{0.92}} \\
    Carlton & \textbf{\underline{-1.38}} & \textbf{\underline{-1.31}} & \textbf{\underline{-1.10}} & \textbf{\underline{-0.93}} & -0.33 & -0.06  \\
    Collingwood & 0.02 & 0.12 & 0.40 & 0.27 & 0.27 & 0.40 \\
    Essendon & -0.55 & -0.42 & 0.00 & 0.00 & -0.24 & -0.38 \\
    Fremantle & -0.24 & \textbf{\underline{-0.85}} & -0.47 & -0.34 & 0.04 & 0.00 \\
    Gold Coast & \textbf{\underline{-1.34}} & \textbf{\underline{-1.54}} & \textbf{\underline{-1.35}} & \textbf{\underline{-1.22}} & \textbf{\underline{-0.82}} & -0.62 \\
    Geelong & \textbf{\underline{0.72}} & \textbf{\underline{0.65}} & \textbf{\underline{0.64}} & \textbf{\underline{0.70}} & \textbf{\underline{1.04}} & \textbf{\underline{0.77}} \\
    Greater Western Sydney & \textbf{\underline{0.63}} & 0.56 & 0.44 & 0.31 & 0.00 & -0.12 \\
    Hawthorn & \textbf{\underline{0.73}} & 0.26 & -0.05 & -0.12 & -0.47 & \textbf{\underline{-0.83}} \\
    Melbourne & 0.03 & -0.23 & -0.07  & 0.34 & 0.42 & \textbf{\underline{0.81}} \\
    North Melbourne & 0.00 & -0.37 & -0.59 &  -0.60 & \textbf{\underline{-1.18}} & \textbf{\underline{-1.90}}\\
    Port Adelaide & 0.26 & 0.00 & 0.44 & \textbf{\underline{0.63}} & 0.58 & \textbf{\underline{0.74}}\\
    Richmond & \textbf{\underline{0.64}} & \textbf{\underline{0.62}} & \textbf{\underline{1.14}} & \textbf{\underline{0.87}} & \textbf{\underline{0.69}} & 0.25 \\
    Saint Kilda & -0.44 & -0.52 & -0.38 & -0.36 & 0.00 & 0.05\\
    Sydney & \textbf{\underline{0.82}} & 0.31 & -0.06 & 0.00 & 0.16 & 0.25 \\
    Western Bulldogs & 0.34 & 0.04 & -0.10 & 0.19 & 0.39 & 0.33 \\
    West Coast & \textbf{\underline{0.92}} & \textbf{\underline{0.61}} & \textbf{\underline{0.68}} & \textbf{\underline{0.62}} & -0.10 & \textbf{\underline{-0.87}} \\
    \bottomrule
    \end{tabular}
}
\end{table}

\begin{table}
    \centering
    \caption{AIC and classification accuracy for the standard Bradley-Terry model, fitted to windows of two to four seasons (2015-2023) and all available data.}
    \label{tab:aic_accuracy_standardw2w4}
        \scalebox{0.7}{
    \begin{tabular}{c c c c c}
    \toprule
    \multicolumn{5}{c}{Window: 2 seasons}\\
    \midrule
 Train Seasons & Test Season & Train AIC & Train Accuracy & Test Accuracy \\
    \cmidrule(lr){1-2} \cmidrule(lr){3-5}
    2015-16 & 2017 & 400.55 & 72.05\% & 58.79\% \\
    2016-17 & 2018 & 418.26 & 67.30\% & 62.70\% \\
    2017-18 & 2019 & 424.11 & 70.96\% & 60.54\% \\
    2018-19 & 2020 & 415.88 & 69.29\% & 56.64\% \\
    2019-20 & 2021 & 397.24 & 66.16\% & 54.14\% \\
    2020-21 & 2022 & 387.31 & 68.52\% & 60.22\% \\
    2021-22 & 2023 & 413.76 & 69.13\% & 63.68\% \\
    2022-23 &  & 422.02 & 70.86\% &  \\
    \midrule
    \multicolumn{5}{c}{Window: 3 seasons}\\
    \midrule
 Train Seasons & Test Season & Train AIC & Train Accuracy & Test Accuracy \\
    \cmidrule(lr){1-2} \cmidrule(lr){3-5}
    2015-17 & 2018 & 540.34 & 67.71\% & 65.76\% \\
    2016-18 & 2019 & 535.43 & 67.82\% & 58.38\% \\
    2017-19 & 2020 & 562.96 & 68.55\% & 56.64\% \\
    2018-20 & 2021 & 522.77 & 65.43\% & 50.83\% \\
    2019-21 & 2022 & 535.23 & 65.16\% & 58.15\% \\
    2020-22 & 2023 & 521.41 & 67.26\% & 65.26\% \\
    2021-23 &  & 548.09 & 69.06\% &  \\
    \midrule
    \multicolumn{5}{c}{Window: 4 seasons}\\
    \midrule
 Train Seasons & Test Season & Train AIC & Train Accuracy & Test Accuracy \\
    \cmidrule(lr){1-2} \cmidrule(lr){3-5}
    2015-18 & 2019 & 632.71 & 66.94\% & 57.30\% \\
    2016-19 & 2020 & 653.18 & 65.99\% & 53.19\% \\
    2017-20 & 2021 & 645.30 & 65.80\% & 57.46\% \\
    2018-21 & 2022 & 634.16 & 65.17\% & 59.67\% \\
    2019-22 & 2023 & 641.68 & 64.44\% & 63.30\% \\
    2020-23 &  & 634.52 & 68.10\% &  \\
    \midrule
    \multicolumn{5}{c}{Window: All Available Data}\\
    \midrule
 Train Seasons & Test Season & Train AIC & Train Accuracy & Test Accuracy \\
    \cmidrule(lr){1-2} \cmidrule(lr){3-5}
    2015-23 &  & 978.62 & 60.98\% & \\

    \bottomrule
    \end{tabular}}
\end{table}

\begin{table}
    \centering
    \caption{Estimated strength and AT\_HOME coefficients with the Contest-Specific Bradley-Terry model, fitted to single seasons (2015-2023) and all available data. Coefficients found to be significant at the 5\% level are in bold and underlined. The coefficient for the reference team is 0.00.}
    \label{tab:coefficients_contest}
    \scalebox{0.7}{
    \begin{tabular}{l *{10}{c}}
    \toprule
    \multicolumn{1}{c}{\textbf{Team}} & \multicolumn{10}{c}{\textbf{Season}} \\
    \cmidrule(lr){2-11}
    & 2015 & 2016 & 2017 & 2018 & 2019 & 2020 & 2021 & 2022 & 2023 & 2015-23 \\
    \midrule
    Adelaide & 0.16 & \textbf{\underline{1.61}} & 1.06 & 0.05 & -0.30 & -1.65 & -0.77 & -1.12 & 0.00 & 0.00 \\
    Brisbane Lions & \textbf{\underline{-2.35}} & \textbf{\underline{-2.22}} & -1.32 & \textbf{\underline{-1.72}} & 0.70 & \textbf{\underline{1.73}} & 0.74 & 0.90 & \textbf{\underline{1.35}} & -0.09 \\
    Carlton & \textbf{\underline{-2.29}} & -0.98 & -1.01 & \textbf{\underline{-3.01}} & -1.02 & -0.31 & -0.57 & 0.06 & 0.63 & \textbf{\underline{-0.63}} \\
    Collingwood & -0.80 & -0.28 & -0.27 & 0.68 & 0.84 & 0.51 & -0.95 & 0.99 & \textbf{\underline{1.73}} & 0.26 \\
    Essendon & \textbf{\underline{-1.58}} & \textbf{\underline{-2.28}} & 0.07 & 0.08 & 0.08 & -0.45 & -0.06 & -1.05 & -0.15 & -0.35 \\
    Fremantle & 1.03 & \textbf{\underline{-1.71}} & -0.58 & -0.93 & -0.45 & -0.39 & -0.13 & 0.79 & -0.37 & -0.17 \\
    Gold Coast & \textbf{\underline{-2.21}} & -1.50 & -1.20 & \textbf{\underline{-2.29}} & \textbf{\underline{-2.08}} & -0.86 & -0.75 & -0.57 & -0.46 & \textbf{\underline{-0.96}} \\
    Geelong & 0.00 & \textbf{\underline{1.79}} & 0.84 & 0.24 & 0.85 & 1.25 & 1.14 & \textbf{\underline{1.67}} & -0.11 & \textbf{\underline{0.67}} \\
    Greater Western Sydney & -0.67 & \textbf{\underline{1.60}} & 0.94 & 0.44 & 0.60 & 0.00 & 0.39 & -1.30 & 0.49 & 0.26 \\
    Hawthorn & 0.94 & \textbf{\underline{1.73}} & -0.06 & 0.41 & 0.08 & -0.90 & -0.66 & -0.96 & -0.95 & 0.01 \\
    Melbourne & -1.36 & -0.03 & 0.18 & 0.43 & -1.36 & 0.25 & \textbf{\underline{2.02}} & 0.80 & 0.76 & 0.20 \\
    North Melbourne & 0.21 & 0.76 & -1.09 & -0.23 & -0.18 & -1.64 & \textbf{\underline{-1.47}} & \textbf{\underline{-3.10}} & \textbf{\underline{-2.26}} & \textbf{\underline{-0.66}} \\
    Port Adelaide & 0.01 & 0.00 & 0.48 & 0.00 & 0.00 & \textbf{\underline{1.75}} & \textbf{\underline{1.34}} & -0.40 & 0.98 & 0.39 \\
    Richmond & 0.30 & -0.58 & 1.13 & 1.35 & 1.23 & \textbf{\underline{1.71}} & -0.13 & 0.12 & -0.21 & \textbf{\underline{0.50}} \\
    Saint Kilda & \textbf{\underline{-1.59}} & 0.47 & 0.09 & \textbf{\underline{-1.87}} & -0.70 & 0.59 & 0.00 & 0.00 & 0.16 & -0.21 \\
    Sydney & 0.50 & \textbf{\underline{1.81}} & 0.64 & 0.45 & -0.72 & -0.93 & 0.75 & 1.03 & 0.20 & 0.37 \\
    Western Bulldogs & 0.05 & \textbf{\underline{1.97}} & 0.00 & -0.94 & 0.16 & 0.45 & 1.22 & 0.04 & 0.13 & 0.29 \\
    West Coast & 1.03 & 1.48 & 0.34 & 1.13 & 0.71 & 0.93 & -0.17 & \textbf{\underline{-2.85}} & \textbf{\underline{-2.23}} & 0.14 \\
    \textbf{Feature (AT\_HOME)} & 0.15 & \textbf{0.63} & \textbf{0.43} & 0.25 & \textbf{0.35} & 0.35 & 0.08 & \textbf{0.62} & \textbf{0.37} & \textbf{0.29} \\
    \bottomrule
    \end{tabular}
}
\end{table}

\begin{table} [H]
    \centering
    \caption{Estimated strength and AT\_HOME coefficients with the Contest-Specific Bradley-Terry model, fitted to windows of two seasons (2015-2023). Coefficients found to be significant at the 5\% level are in bold and underlined. The coefficient for the reference team is 0.00.}
    \label{tab:coefficients_contestw2}
    \scalebox{0.7}{
    \begin{tabular}{l *{8}{c}}
    \toprule
    \multicolumn{1}{c}{\textbf{Team}} & \multicolumn{8}{c}{\textbf{Season}} \\
    \cmidrule(lr){2-9}
    & 2015-16 & 2016-17 & 2017-18 & 2018-19 & 2019-20 & 2020-21 & 2021-22 & 2022-23 \\
    \midrule
    Adelaide & 0.88 & 0.91 & 0.39 & -0.03 & -0.60 & -1.13 & -0.84 & -0.52 \\
    Brisbane Lions & \textbf{\underline{-1.99}} & \textbf{\underline{-1.86}} & \textbf{\underline{-1.67}} & -0.22 & \textbf{\underline{1.18}} & \textbf{\underline{0.99}} & 0.82 & \textbf{\underline{1.00}} \\
    Carlton & \textbf{\underline{-1.44}} & \textbf{\underline{-1.15}} & \textbf{\underline{-1.93}} & \textbf{\underline{-1.61}} & -0.52 & -0.48 & -0.21 & 0.32  \\
    Collingwood & -0.41 & -0.49 & 0.04 & 0.85 & 0.78 & -0.34 & -0.02 & \textbf{\underline{1.18}} \\
    Essendon & \textbf{\underline{-1.65}} & \textbf{\underline{-1.00}} & -0.14 & 0.21 & 0.00 & -0.27 & -0.43 & -0.56 \\
    Fremantle & 0.00 & \textbf{\underline{-1.19}} & -0.88 & -0.53 & -0.25 & -0.27 & 0.34 & 0.22 \\
    Gold Coast & \textbf{\underline{-1.54}} & \textbf{\underline{-1.38}} & \textbf{\underline{-1.78}} & \textbf{\underline{-1.91}} & \textbf{\underline{-1.31}} & -0.82 & -0.58 & -0.52 \\
    Geelong & 0.84 & \textbf{\underline{0.91}} & 0.35 & 0.61 & \textbf{\underline{1.02}} & \textbf{\underline{0.98}} & \textbf{\underline{1.31}} & 0.67 \\
    Greater Western Sydney & 0.51 & 0.87 & 0.47 & 0.59 & 0.46 & 0.12 & -0.28 & -0.31 \\
    Hawthorn & \textbf{\underline{1.25}} & 0.44 & 0.00 & 0.33 & -0.18 & -0.80 & -0.72 & \textbf{\underline{-0.96}} \\
    Melbourne & -0.58 & -0.14 & 0.14 & -0.22 & -0.46 & \textbf{\underline{1.05}} & \textbf{\underline{1.29}} & 0.73 \\
    North Melbourne & 0.40 & -0.52 & -0.82 & -0.07 & -0.56 &  \textbf{\underline{-1.55}} & \textbf{\underline{-1.92}} & \textbf{\underline{-2.59}}\\
    Port Adelaide & 0.05 & 0.01 & 0.08 & 0.10 & 0.78 & \textbf{\underline{1.32}} & 0.53 & 0.31\\
    Richmond & -0.01 & 0.17 & \textbf{\underline{0.95}} & \textbf{\underline{1.33}} & \textbf{\underline{1.44}} & 0.57 & 0.07 & -0.01 \\
    Saint Kilda & -0.47 & 0.00 & -0.88 & \textbf{\underline{-1.02}} & 0.03 & 0.19 & 0.00 & 0.00\\
    Sydney & \textbf{\underline{1.06}} & 0.86 & 0.30 & 0.00 & -0.62 & 0.00 & 0.86 & 0.56 \\
    Western Bulldogs & \textbf{\underline{0.94}} & 0.58 & -0.60 & -0.28 & 0.37 & 0.75 & 0.61 & 0.05 \\
    West Coast & \textbf{\underline{1.18}} & 0.48 & 0.45 & \textbf{\underline{1.00}} & 0.88 & 0.21 & \textbf{\underline{-1.7}} & \textbf{\underline{-2.51}} \\
    \textbf{Feature (AT\_HOME)}  & \textbf{\underline{0.35}} & \textbf{\underline{0.47}}  & \textbf{\underline{0.35}} & \textbf{\underline{0.29}} & \textbf{\underline{0.34}} & 0.18 & \textbf{\underline{0.29}} & \textbf{\underline{0.47}} \\
    \bottomrule
    \end{tabular}
}
\end{table}

\begin{table} [H]
    \centering
    \caption{Estimated strength and AT\_HOME coefficients with the Contest-Specific Bradley-Terry model, fitted to windows of three seasons (2015-2023). Coefficients found to be significant at the 5\% level are in bold and underlined. The coefficient for the reference team is 0.00.}
    \label{tab:coefficients_contestw3}
    \scalebox{0.7}{
    \begin{tabular}{l *{7}{c}}
    \toprule
    \multicolumn{1}{c}{\textbf{Team}} & \multicolumn{7}{c}{\textbf{Season}} \\
    \cmidrule(lr){2-8}
    & 2015-17 & 2016-18 & 2017-19 & 2018-20 & 2019-21 & 2020-22 & 2021-23\\
    \midrule
    Adelaide & \textbf{\underline{0.94}} & 0.58 & 0.16 & -0.40 & -0.66 & \textbf{\textbf{\underline{-1.12}}} & -0.58 \\
    Brisbane Lions & \textbf{\underline{-1.63}} & \textbf{\underline{-1.78}} & \textbf{\underline{-0.73}} & 0.21 & \textbf{\underline{0.91}} & \textbf{\underline{0.83}} & \textbf{\underline{0.88}} \\
    Carlton & \textbf{\underline{-1.16}} & \textbf{\underline{-1.56}} & \textbf{\underline{-1.54}} & \textbf{\underline{-1.17}} & -0.53 & -0.37 & 0.07  \\
    Collingwood & -0.26 & -0.12 & 0.31 & 0.67 & 0.16 & -0.06 & 0.49 \\
    Essendon & \textbf{\underline{-0.84}} & -0.67 & -0.04 & 0.00 & -0.06 & -0.57 & -0.33 \\
    Fremantle  & -0.12 & \textbf{\underline{-1.09}} & -0.69 & -0.48 & -0.22 & 0.00 & 0.13 \\
    Gold Coast & \textbf{\underline{-1.25}} & \textbf{\underline{-1.58}} & \textbf{\underline{-1.76}} & \textbf{\underline{-1.53}} & \textbf{\underline{-1.08}} & -0.76 & -0.55 \\
    Geelong & \textbf{\underline{0.87}} & 0.62 & 0.48 & 0.64 & \textbf{\underline{0.91}} & \textbf{\underline{1.04}} & \textbf{\underline{0.79}} \\
    Greater Western Sydney  & 0.70 & 0.65 & 0.48 & 0.36 & 0.32 & -0.37 & -0.05 \\
    Hawthorn & \textbf{\underline{0.85}} & 0.38 & 0.00 & 0.00 & -0.38 & \textbf{\underline{-0.88}} & \textbf{\underline{-0.79}} \\
    Melbourne & -0.22 & 0.07 & -0.27 & -0.14  & 0.29 & \textbf{\underline{0.81}} & \textbf{\underline{1.02}} \\
    North Melbourne & 0.00 & -0.43 & -0.59 & -0.43 &  \textbf{\underline{-0.86}} & \textbf{\underline{-1.95}} & \textbf{\underline{-1.99}}\\
    Port Adelaide & 0.26 & 0.00 & 0.05 & 0.43 & \textbf{\underline{0.85}} & 0.63 & 0.62\\
    Richmond & 0.43 & 0.46 & \textbf{\underline{1.02}} & \textbf{\underline{1.25}} & \textbf{\underline{0.84}} & 0.35 & 0.00 \\
    Saint Kilda & -0.16 & -0.52 & \textbf{\underline{-0.78}} & -0.54 & 0.00 & 0.00 & 0.00\\
    Sydney & \textbf{\underline{0.95}} & 0.64 & 0.00 & -0.28 & -0.14 & 0.24 & 0.60 \\
    Western Bulldogs & 0.69 & 0.09 & -0.36 & -0.12 & 0.54 & 0.38 & 0.42 \\
    West Coast & \textbf{\underline{0.87}} & 0.61 & 0.54 & \textbf{\underline{0.88}} & 0.45 & -0.63 & \textbf{\underline{-1.39}} \\
     \textbf{Feature (AT\_HOME)}  & \textbf{\underline{0.35}} & \textbf{\underline{0.39}} & \textbf{\underline{0.34}} & \textbf{\underline{0.30}} & \textbf{\underline{0.24}} & \textbf{\underline{0.30}} & \textbf{\underline{0.32}} \\
    \bottomrule
    \end{tabular}
}
\end{table}

\begin{table} [H]
    \centering
    \caption{Estimated strength and AT\_HOME coefficients with the Contest-Specific Bradley-Terry model, fitted to windows of four seasons (2015-2023). Coefficients found to be significant at the 5\% level are in bold and underlined. The coefficient for the reference team is 0.00.}
    \label{tab:coefficients_contestw4}
    \scalebox{0.7}{
    \begin{tabular}{l *{6}{c}}
    \toprule
    \multicolumn{1}{c}{\textbf{Team}} & \multicolumn{6}{c}{\textbf{Season}} \\
    \cmidrule(lr){2-7}
    & 2015-18 & 2016-19 & 2017-20 & 2018-21 & 2019-22 & 2020-23\\
    \midrule
    Adelaide & \textbf{\underline{0.73}} & 0.36 & -0.02 & -0.43 & \textbf{\textbf{\underline{-0.69}}} & \textbf{\underline{-0.75}} \\
    Brisbane Lions & \textbf{\underline{-1.58}} & \textbf{\underline{-0.95}} & -0.15 & 0.33 & \textbf{\underline{0.86}} & \textbf{\underline{0.95}} \\
    Carlton & \textbf{\underline{-1.41}} & \textbf{\underline{-1.33}} & \textbf{\underline{-1.13}} & \textbf{\underline{-0.94}} & -0.35 & -0.04  \\
    Collingwood & 0.02 & 0.14 & 0.43 & 0.29 & 0.28 & 0.43 \\
    Essendon & -0.57 & -0.43 & 0.00 & 0.01 & -0.25 & -0.37 \\
    Fremantle &  -0.26 & \textbf{\underline{-0.87}} & -0.50 & -0.35 & 0.02 & 0.00 \\
    Gold Coast & \textbf{\underline{-1.39}} & \textbf{\underline{-1.59}} & \textbf{\underline{-1.40}} & \textbf{\underline{-1.24}} & \textbf{\underline{-0.86}} & -0.62 \\
    Geelong & \textbf{\underline{0.71}} & \textbf{\underline{0.66}} & \textbf{\underline{0.65}} & \textbf{\underline{0.72}} & \textbf{\underline{1.04}} & \textbf{\underline{0.80}} \\
    Greater Western Sydney & \textbf{\underline{0.64}} & 0.60 & 0.47 & 0.33 & 0.00 & -0.09 \\
    Hawthorn & \textbf{\underline{0.75}} & 0.28 & -0.05 & -0.12 & -0.49 & \textbf{\underline{-0.82}} \\
    Melbourne & 0.02 & -0.23 & -0.07  & 0.34 & 0.40 & \textbf{\underline{0.81}} \\
    North Melbourne & 0.00 & -0.37 & -0.60 &  -0.60 & \textbf{\underline{-1.20}} & \textbf{\underline{-1.90}}\\
    Port Adelaide & 0.24 & 0.00 & 0.43 & 0.62 & 0.56 & \textbf{\underline{0.75}}\\
    Richmond & \textbf{\underline{0.65}} & \textbf{\underline{0.63}} & \textbf{\underline{1.16}} & \textbf{\underline{0.88}} & \textbf{\underline{0.69}} & 0.28 \\
    Saint Kilda & -0.45 & -0.53 & -0.38 & -0.36 & -0.01 & 0.08\\
    Sydney & \textbf{\underline{0.82}} & 0.32 & -0.06 & 0.00 & 0.15 & 0.29 \\
    Western Bulldogs & 0.36 & 0.07 & -0.09 & 0.21 & 0.41 & 0.38 \\
    West Coast & \textbf{\underline{0.91}} & \textbf{\underline{0.63}} & \textbf{\underline{0.68}} & \textbf{\underline{0.62}} & -0.13 & \textbf{\underline{-0.88}} \\
    \textbf{Feature (AT\_HOME)}  & \textbf{\underline{0.32}} & \textbf{\underline{0.36}} & \textbf{\underline{0.33}} & \textbf{\underline{0.24}} & \textbf{\underline{0.30}} & \textbf{\underline{0.31}} \\
    \bottomrule
    \end{tabular}
}
\end{table}

\begin{table}[H]
    \centering
    \caption{AIC and classification accuracy for the Contest-Specific Bradley-Terry model, fitted to windows of two to four seasons (2015-2023) and all available data.}
    \label{tab:aic_accuracy_contestw2w4}
        \scalebox{0.7}{
    \begin{tabular}{c c c c c}
    \toprule
    \multicolumn{5}{c}{Window: 2 seasons}\\
    \midrule
 Train Seasons & Test Season & Train AIC & Train Accuracy & Test Accuracy \\
    \cmidrule(lr){1-2} \cmidrule(lr){3-5}
    2015-16 & 2017 & 393.84 & 72.33\% & 59.89\% \\
    2016-17 & 2018 & 403.27 & 68.66\% & 64.32\% \\
    2017-18 & 2019 & 416.55 & 69.86\% & 60.54\% \\
    2018-19 & 2020 & 411.12 & 69.57\% & 60.84\% \\
    2019-20 & 2021 & 390.83 & 67.68\% & 56.35\% \\
    2020-21 & 2022 & 386.91 & 68.52\% & 61.88\% \\
    2021-22 & 2023 & 408.92 & 68.85\% & 65.26\% \\
    2022-23 &  & 407.56 & 74.33\% &  \\
    \midrule
    \multicolumn{5}{c}{Window: 3 seasons}\\
    \midrule
 Train Seasons & Test Season & Train AIC & Train Accuracy & Test Accuracy \\
    \cmidrule(lr){1-2} \cmidrule(lr){3-5}
    2015-17 & 2018 & 527.27 & 68.26\% & 65.76\% \\
    2016-18 & 2019 & 519.56 & 68.73\% & 62.16\% \\
    2017-19 & 2020 & 550.78 & 68.18\% & 55.24\% \\
    2018-20 & 2021 & 514.49 & 65.62\% & 55.80\% \\
    2019-21 & 2022 & 530.17 & 66.14\% & 61.96\% \\
    2020-22 & 2023 & 513.40 & 67.26\% & 67.37\% \\
    2021-23 &  & 537.55 & 70.50\% &  \\
    \midrule
    \multicolumn{5}{c}{Window: 4 seasons}\\
    \midrule
 Train Seasons & Test Season & Train AIC & Train Accuracy & Test Accuracy \\
    \cmidrule(lr){1-2} \cmidrule(lr){3-5}
    2015-18 & 2019 & 618.02 & 68.59\% & 61.92\% \\
    2016-19 & 2020 & 632.80 & 67.58\% & 56.74\% \\
    2017-20 & 2021 & 629.43 & 65.51\% & 56.35\% \\
    2018-21 & 2022 & 626.85 & 65.61\% & 61.33\% \\
    2019-22 & 2023 & 628.95 & 65.17\% & 64.36\% \\
    2020-23 &  & 620.77 & 65.95\% &  \\
    \midrule
    \multicolumn{5}{c}{Window: All Available Data}\\
    \midrule
 Train Seasons & Test Season & Train AIC & Train Accuracy & Test Accuracy \\
    \cmidrule(lr){1-2} \cmidrule(lr){3-5}
    2015-23 &  & 945.57 & 61.92\% & \\

    \bottomrule
    \end{tabular}}
\end{table}

\begin{table}[H]
    \centering
    \caption{Estimated coefficients for the time-variant Bradley-Terry fitted to windows of one season and all available data with PIs encoded as cumulatives over the season (2015-2023). Significant coefficients are in bold and underlined. Significant coefficients at the individual level only are in italics.
}
    \label{tab:bt_expansion_coefficientsseasonw1}
    \scalebox{0.53}{
    \begin{tabular}{lcccccccccc}
    \toprule
   &  \multicolumn{10}{c}{\textbf{Season}} \\
       \cmidrule{2-11}
     \textbf{FEATURE} & \textbf{2015} & \textbf{2016} & \textbf{2017} & \textbf{2018} & \textbf{2019} & \textbf{2020} & \textbf{2021} & \textbf{2022}& \textbf{2023}& \textbf{2015-23}\\
    \midrule
   & \multicolumn{10}{c}{MATCH DIFFICULTY}\\
    \midrule 
    AT\_HOME &  & \textbf{\underline{0.599}} & \textit{0.392} &  & \textit{0.303} & & &\textit{0.521} & \textbf{\underline{0.551}} &\textbf{\underline{0.182}}  \\
  HOMEGROUND & \textit{0.381} & \textit{0.316} &&& \textbf{\underline{0.653}}&& &  \textbf{\underline{0.746}} &  \textit{0.408} & \textit{0.363} \\ 
  INTERSTATE & \textbf{\underline{-0.459}} & \textit{-0.443} & \textbf{\underline{-0.506}}&&\textit{-0.530} & \textbf{\underline{-0.645}}& & \textit{-0.515} &  \textit{-0.437} & \textbf{\underline{-0.268}}  \\
      \midrule
   & \multicolumn{10}{c}{FORM}\\
    \midrule 
CONSECUTIVE\_LOSSES  &  &  & && && \textbf{\underline{0.344}}&& \textit{-0.164}&\textit{-0.139}  \\ 
CONSECUTIVE\_WINS &  &  & && && && &\textit{0.106}\\
LADDER\_POSITION\_DIFF & \textit{0.039}&\textit{0.061}&\textit{0.038}&\textit{0.030}&&\textit{0.029}&&\textit{0.050}&&\textit{0.044}\\
LADDERLY\_POSITION\_DIFF & \textbf{\underline{0.056}}&\textbf{\underline{0.043}}&\textit{0.034}&\textbf{\underline{0.077}}&\textit{0.037}&\textbf{\underline{0.043}}&\textit{0.036}&&\textbf{\underline{0.027}}&\textbf{\underline{0.025}}\\
LG\_WON  &  &  & && && && \textbf{\underline{0.787}}&\textit{0.278}\\
PERCENTAGE\_DIFF &  &  & \textit{0.009}&\textit{0.007}& &&\textbf{\underline{0.010}} &\textit{0.012}&&\textit{0.009}\\
POINTSAGAINST\_DIFF & && & &  &  & \textit{0.002} & \textit{0.002} & \textit{0.001} & \textbf{\underline{0.001}} \\
POINTSFOR\_DIFF & \textit{0.001}&& \textbf{\underline{0.003}} &  \textbf{\underline{0.003}}& \textit{0.003} &  \textit{0.003} & & \textbf{\underline{-0.005}} & \textit{0.001} & \textbf{\underline{0.001}} \\
WINS\_CUMULATIVE\_DIFF & && \textbf{\underline{-0.268}} &  \textbf{\underline{-0.148}}& &  \textit{0.130} & & \textit{0.133}&  & \textit{0.100} \\
   \midrule
   & \multicolumn{10}{c}{PIs}\\
    \midrule 
BOUNCES\_CSUM\_DIFF &  &  & && && && &\textit{0.002}\\
CLANGERS\_CSUM\_DIFF    \\
CLEARANCES\_CENTRE\_CSUM\_DIFF & \textit{0.012} &  & && && && &\textit{0.006} \\
CLEARANCES\_CSUM\_DIFF &  &\textit{0.006}  & && \textit{-0.004}&& && &\textit{0.003} \\
CLEARANCES\_STOPPAGE\_CSUM\_DIFF & & & && \textit{-0.005}&& && &\textit{0.002}\\
CONTEST\_DEFENSIVE\_LOSS\_CSUM\_DIFF & & & && \textbf{\underline{0.022}}&& && &\textit{0.009}\\
CONTEST\_DEFENSIVE\_LOSS\_RATE\_CSUM\_DIFF & & & && \textit{-0.031}&& && &\textit{-0.014}\\
CONTEST\_OFFENSIVE\_WIN\_CSUM\_DIFF & & & \textit{0.013}&& && && &\textit{0.006}\\
CONTEST\_OFFENSIVE\_WIN\_RATE\_CSUM\_DIFF \\
DISPOSALS\_CSUM\_DIFF &&&&&&&&&&\textbf{\underline{0.001}}\\
DISPOSALS\_EFFECTIVE\_CSUM\_DIFF &&&&&&&&&&\textit{0.000}\\
DISPOSALS\_EFFICIENCY\_CSUM\_DIFF \\
FREES\_AGAINST\_CSUM\_DIFF & & & \textbf{\underline{-0.011}}\\
GETS\_GROUNDBALL\_CSUM\_DIFF & & & && \textit{-0.002}&&\textbf{\underline{0.003}}&& &\textbf{\underline{-0.001}}\\
GETS\_GROUNDBALL50\_CSUM\_DIFF & & &\textit{0.006} && &&\textit{0.008}&& &\textbf{\underline{0.004}}\\
GOALS\_ACCURACY\_CSUM\_DIFF &&&&&&&&&&\textit{0.009} \\
GOALS\_SHOTS\_CSUM\_DIFF & & &\textbf{\underline{-0.029}}&\textit{0.010}&\textbf{\underline{0.011}} &\textbf{\underline{0.027}}&&\textit{0.009}& &\textit{0.008}\\
HANDBALLS\_CSUM\_DIFF \\
HITOUTS\_ADVANTAGE\_CSUM\_DIFF &&&&&&&&&&\textit{0.002}\\
HITOUTS\_ADVANTAGE\_RATE\_CSUM\_DIFF &&&&&&&&&&\textit{0.010}\\
HITOUTS\_WIN\_RATE\_CSUM\_DIFF  & & &\textbf{\underline{-0.003}}&\\
INSIDE50\_CSUM\_DIFF & &\textbf{\underline{0.005}}&\textit{0.003}&\textit{0.004}&&\textbf{\underline{0.014}}&&\textit{0.006}& \textbf{\underline{0.005}} &\textit{0.005} \\
INTERCEPTS\_CSUM\_DIFF &&&&&&&&&&\textit{0.002}\\
KICK2HANDBALL\_CSUM\_DIFF &\textit{-1.535}&&&&&&&&&\textbf{\underline{0.371}}\\
KICKS\_CSUM\_DIFF &&&&&&&&&&\textbf{\underline{-0.001}}\\
KICKS\_EFFECTIVE\_CSUM\_DIFF  &&&&&&\textit{0.002}&&&&\textit{0.001}\\
KICKS\_EFFICIENCY\_CSUM\_DIFF &&&&&\textit{0.060}\\
MARKS\_CONTESTED\_CSUM\_DIFF  &&\textit{0.011}&&\textit{0.013}&&&&& &\textit{0.008}\\
MARKS\_CSUM\_DIFF  & &&&&&\textit{0.003}&&& &\textit{0.001}\\
MARKS\_INSIDE50\_CSUM\_DIFF &&&\textit{0.008}&&&\textit{0.021}&&\textit{0.014}& &\textit{0.008}\\
MARKS\_INTERCEPT\_CSUM\_DIFF &&&&\textit{0.009}&&&&& &\textit{0.005}\\
MARKS\_ONLEAD\_CSUM\_DIFF &&&&&&&&& &\textit{0.003}\\
METRES\_GAINED\_L4\_CSUM\_DIFF &&&&\textit{0.000}&&&&& &\textit{0.000}\\
ONE\_PERCENTERS\_CSUM\_DIFF &&&&&&&&&&\textit{0.001}\\
POSSESSIONS\_CONTESTED\_CSUM\_DIFF &&&&&\textbf{\underline{-0.002}}&&\textit{0.003}&&&\textit{0.002}\\
POSSESSIONS\_CONTESTED\_RATE\_CSUM\_DIFF\\
POSSESSIONS\_CSUM\_DIFF &&&&&&&&&&\textit{0.000}\\
POSSESSIONS\_UNCONTESTED\_CSUM\_DIFF&&&&&&&&&&\textit{0.000}\\
PRESSURE\_CSUM\_DIFF\\
PRESSURE\_DEFENSEHALF\_CSUM\_DIFF &&\textbf{\underline{-0.002}}&&\textit{-0.002}&&&&& &\textit{-0.001}\\
REBOUND\_INSIDE50S\_CSUM\_DIFF &\textbf{\underline{-0.008}}&&&&&\textit{-0.008}&&\textit{-0.010}& &\textit{-0.004}\\
SCORE\_LAUNCHES\_CSUM\_DIFF &&&\textbf{\underline{0.039}}&\textit{0.009}&\textit{0.009}&\textbf{\underline{-0.037}}&&\textbf{\underline{0.030}}&\textit{0.005}&\textit{0.008}\\
SPOILS\_CSUM\_DIFF\\
TACKLES\_CSUM\_DIFF &&&&&&&\textit{0.003}\\
TACKLES\_INSIDE50\_L4\_CSUM\_DIFF &&&&&&&\textit{0.009}&&&\textit{0.006}\\
TURNOVERS\_L4\_CSUM\_DIFF &&&&&&\textbf{\underline{0.009}}\\
\bottomrule
    \end{tabular}}
    \end{table}

\begin{table}[H]
    \centering
    \caption{Estimated coefficients for the time-variant Bradley-Terry fitted to windows of two seasons with PIs encoded as cumulatives over the previous four games (2015-2023). Significant coefficients are in bold and underlined. Significant coefficients at the individual level only are in italics.
}
    \label{tab:bt_expansion_coefficientsl4w2}
    \scalebox{0.53}{
    \begin{tabular}{lcccccccccc}
    \toprule
   &  \multicolumn{8}{c}{\textbf{Season}} \\
       \cmidrule{2-9}
     \textbf{FEATURE} & \textbf{2015-16} & \textbf{2016-17} & \textbf{2017-18} & \textbf{2018-19} & \textbf{2019-20} & \textbf{2020-21} & \textbf{2021-22} & \textbf{2022-23}\\
    \midrule
   & \multicolumn{8}{c}{MATCH DIFFICULTY}\\
    \midrule 
    AT\_HOME & \textit{0.005} & \textbf{\underline{0.465}} & \textbf{\underline{0.324}} & \textit{0.268} & \textit{0.312} & & \textit{0.279} & \textit{0.431} \\
  HOMEGROUND &  \textit{0.003} & \textit{0.316} & \textit{0.257} &  \textbf{\underline{0.464}} &  \textbf{\underline{0.610}} & & \textbf{\underline{0.406}} & \textbf{\underline{0.563}} \\ 
  INTERSTATE & \textbf{\underline{-0.497}} & \textit{-0.463} & \textit{-0.348} &\textit{-0.359} & \textit{-0.583} & \textbf{\underline{-0.351}} &  \textit{-0.358} & \textit{-0.488}  \\
      \midrule
   & \multicolumn{8}{c}{FORM}\\
    \midrule 
CONSECUTIVE\_LOSSES  & & \textit{-0.092} &&& \textit{-0.124} \\ 
CONSECUTIVE\_WINS &  &&&&&& &\textit{0.111}\\
L4G\_WINS  & &\textit{0.161}\\
LADDER\_POSITION\_DIFF & \textit{0.038}&\textit{0.035}&\textit{0.023}&\textit{0.019}&&&\textit{0.032}&\textit{0.022}\\
LADDERLY\_POSITION\_DIFF & \textbf{\underline{0.039}}&&\textbf{\underline{0.030}}&\textbf{\underline{0.033}}&\textbf{\underline{0.029}}\\
LG\_WON &&&&&&&&& \\
PERCENTAGE\_DIFF & \textit{0.005} & \textbf{\underline{0.005}} & \textit{0.007} & \textit{0.005} & & & \textit{0.009} & \textit{0.005}\\
POINTSAGAINST\_DIFF & \textbf{\underline{0.001}} & \textbf{\underline{0.001}} & \textbf{\underline{0.001}} & \textit{0.001} & \textit{0.001} & \textbf{\underline{0.001}} & \textbf{\underline{0.002}} & \textbf{\underline{0.001}} \\
POINTSFOR\_DIFF & \textbf{\underline{0.001}} & \textit{0.001} & \textbf{\underline{0.001}} &  \textbf{\underline{0.002}} & \textbf{\underline{0.002}} & & \textit{0.001} & \textit{0.001} \\
      \midrule
   & \multicolumn{8}{c}{PIs}\\
    \midrule 
BOUNCES\_L4\_CSUM\_DIFF \\
CLANGERS\_L4\_CSUM\_DIFF  \\
CLEARANCES\_CENTRE\_L4\_CSUM\_DIFF &\textit{0.012} &\textit{0.012} \\
CLEARANCES\_L4\_CSUM\_DIFF &&&&&&\textit{0.010} \\
CLEARANCES\_STOPPAGE\_L4\_CSUM\_DIFF &&&&&&\textit{0.011} \\
CONTEST\_DEFENSIVE\_LOSS\_L4\_CSUM\_DIFF &&&&&\textbf{\underline{0.024}}\\
CONTEST\_DEFENSIVE\_LOSS\_RATE\_L4\_CSUM\_DIFF &&&& \textit{-0.014} &\textit{-0.014} \\
CONTEST\_OFFENSIVE\_WIN\_L4\_CSUM\_DIFF \\
CONTEST\_OFFENSIVE\_WIN\_RATE\_L4\_CSUM\_DIFF \\
DISPOSALS\_EFFECTIVE\_L4\_CSUM\_DIFF \\
DISPOSALS\_EFFICIENCY\_L4\_CSUM\_DIFF &&&&&&&\textit{-0.043}\\
DISPOSALS\_L4\_CSUM\_DIFF \\
FREES\_AGAINST\_L4\_CSUM\_DIFF \\
GETS\_GROUNDBALL\_L4\_CSUM\_DIFF &&&&&&\textit{0.005}&\textit{0.005} \\
GETS\_GROUNDBALL50\_L4\_CSUM\_DIFF &&&&&&\textbf{\underline{0.014}}&\textbf{\underline{0.014}}&\textit{0.009}\\
GOALS\_ACCURACY\_L4\_CSUM\_DIFF \\
GOALS\_SHOTS\_L4\_CSUM\_DIFF &&&\textit{0.008}&\textbf{\underline{0.009}}&\textbf{\underline{0.011}}&&&\textit{0.008}\\
HANDBALLS\_L4\_CSUM\_DIFF \\
HITOUTS\_ADVANTAGE\_L4\_CSUM\_DIFF &&&\textit{0.008}&\textit{0.009}\\
HITOUTS\_ADVANTAGE\_RATE\_L4\_CSUM\_DIFF  \\
HITOUTS\_WIN\_RATE\_L4\_CSUM\_DIFF &\textit{0.004} &&\textit{0.003}  \\
INSIDE50\_L4\_CSUM\_DIFF &\textit{0.007}&\textit{0.006}&\textit{0.006}&\textit{0.008}&&\textit{0.006}&\textit{0.005}&\textit{0.007} \\
INTERCEPTS\_L4\_CSUM\_DIFF  &&&&\textit{0.004}&&&\textit{0.005}&\textit{0.004}\\
KICK2HANDBALL\_L4\_CSUM\_DIFF  \\
KICKS\_EFFECTIVE\_L4\_CSUM\_DIFF  \\
KICKS\_EFFICIENCY\_L4\_CSUM\_DIFF\\
KICKS\_L4\_CSUM\_DIFF \\
MARKS\_CONTESTED\_L4\_CSUM\_DIFF&&&\textit{0.011}\\
MARKS\_INSIDE50\_L4\_CSUM\_DIFF \\
MARKS\_INTERCEPT\_L4\_CSUM\_DIFF&&&\textbf{\underline{0.009}}\\
MARKS\_L4\_CSUM\_DIFF\\
MARKS\_ONLEAD\_L4\_CSUM\_DIFF\\
METRES\_GAINED\_L4\_CSUM\_DIFF&&\textit{0.000}&\textit{0.000}&\textit{0.000}\\
ONE\_PERCENTERS\_L4\_CSUM\_DIFF \\
POSSESSIONS\_CONTESTED\_L4\_CSUM\_DIFF &&&&\textit{0.002}&&\textit{0.005}&\textit{0.001}\\
POSSESSIONS\_CONTESTED\_RATE\_L4\_CSUM\_DIFF &&&&&&&\textit{0.030}\\
POSSESSIONS\_L4\_CSUM\_DIFF \\
POSSESSIONS\_UNCONTESTED\_L4\_CSUM\_DIFF\\
PRESSURE\_DEFENSEHALF\_L4\_CSUM\_DIFF& \textbf{\underline{-0.002}}\\
PRESSURE\_L4\_CSUM\_DIFF\\
REBOUND\_INSIDE50S\_L4\_CSUM\_DIFF&\textit{-0.009}\\
SCORE\_LAUNCHES\_L4\_CSUM\_DIFF&& &\textit{0.008}&\textit{0.147}&\textit{0.012}&&\textit{0.007}&\textbf{\underline{0.012}}\\
SPOILS\_L4\_CSUM\_DIFF\\
TACKLES\_INSIDE50\_L4\_CSUM\_DIFF&& &&&\textit{0.009}&&\textit{0.010}\\
TACKLES\_L4\_CSUM\_DIFF\\
TURNOVERS\_L4\_CSUM\_DIFF\\
\bottomrule
    \end{tabular}}
    \end{table}

\begin{table}[H]
    \centering
    \caption{Estimated coefficients for the time-variant Bradley-Terry fitted to windows of two seasons with PIs encoded as cumulatives over the season (2015-2023). Significant coefficients are in bold and underlined. Significant coefficients at the individual level only are in italics.
}
    \label{tab:bt_expansion_coefficientsseasonw2}
    \scalebox{0.53}{
    \begin{tabular}{lcccccccccc}
    \toprule
   &  \multicolumn{8}{c}{\textbf{Season}} \\
       \cmidrule{2-9}
     \textbf{FEATURE} & \textbf{2015-16} & \textbf{2016-17} & \textbf{2017-18} & \textbf{2018-19} & \textbf{2019-20} & \textbf{2020-21} & \textbf{2021-22} & \textbf{2022-23}\\
    \midrule
   & \multicolumn{8}{c}{MATCH DIFFICULTY}\\
    \midrule 
    AT\_HOME & \textit{0.316} & \textbf{\underline{0.475}} & \textbf{\underline{0.324}} & \textit{0.268} & \textit{0.312} & & \textit{0.279} & \textit{0.007} \\
  HOMEGROUND &  \textbf{\underline{0.458}} & \textit{0.316} & \textit{0.257} &  \textbf{\underline{0.469}} &  \textbf{\underline{0.688}} & & \textbf{\underline{0.438}} & \textbf{\underline{0.616}} \\ 
  INTERSTATE & \textit{-0.457} & \textit{-0.463} & \textit{-0.348} &\textit{-0.359} & \textit{-0.583} & \textbf{\underline{-0.337}} &  \textit{-0.358} & \textit{-0.488}  \\
      \midrule
   & \multicolumn{8}{c}{FORM}\\
    \midrule 
CONSECUTIVE\_LOSSES  & & \textit{-0.092} &&& \textit{-0.124} \\ 
CONSECUTIVE\_WINS &  &&&&&& &\textbf{\underline{0.133}}\\
LADDER\_POSITION\_DIFF & \textit{0.038}&\textbf{\underline{0.029}}&\textit{0.023}&\textit{0.019}&&&\textit{0.032}&\textit{0.022}\\
LADDERLY\_POSITION\_DIFF & \textbf{\underline{0.047}}&&\textbf{\underline{0.030}}&\textbf{\underline{0.038}}&\textbf{\underline{0.026}}\\
LG\_WON \\
PERCENTAGE\_DIFF & \textit{0.005} & \textit{0.007} & \textit{0.007} & \textit{0.005} & & & \textbf{\underline{0.006}} & \textit{0.005}\\
POINTSAGAINST\_DIFF & \textit{0.001} & \textit{0.002} & \textbf{\underline{0.001}} & \textit{0.001} & \textit{0.001} & \textit{0.001} & \textit{0.002} & \textbf{\underline{0.002}} \\
POINTSFOR\_DIFF & \textit{0.001} & \textit{0.001} & \textbf{\underline{0.001}} &  \textit{0.002} & \textit{0.002} & & \textit{0.001} & \textit{0.001} \\
WINS\_CUMULATIVE\_DIFF & \textit{0.058} & \textit{0.067} & \textit{0.052}&  \textit{0.060} & \textbf{\underline{-0.161}} & & \textit{0.070} & \textbf{\underline{-0.104}} \\
      \midrule
   & \multicolumn{8}{c}{PIs}\\
    \midrule 
BOUNCES\_CSUM\_DIFF \\
CLANGERS\_CSUM\_DIFF   &  &  & &   &  & &  \textbf{\underline{-0.004}} \\
CLEARANCES\_CENTRE\_CSUM\_DIFF &\textit{0.009} &\textit{0.007} \\
CLEARANCES\_CSUM\_DIFF & \textit{0.004} &  & &   &    \textbf{\underline{-0.005}} \\
CLEARANCES\_STOPPAGE\_CSUM\_DIFF &&&&&\textit{-0.006} \\
CONTEST\_DEFENSIVE\_LOSS\_CSUM\_DIFF &  &  &\textit{0.013} && \textbf{\underline{0.031}}\\
CONTEST\_DEFENSIVE\_LOSS\_RATE\_CSUM\_DIFF &  &  & &\textit{-0.018}& \textit{-0.021}&\textit{-0.022}\\
CONTEST\_OFFENSIVE\_WIN\_CSUM\_DIFF \\
CONTEST\_OFFENSIVE\_WIN\_RATE\_CSUM\_DIFF \\
DISPOSALS\_CSUM\_DIFF & \textit{0.000}&  & && &\textit{0.001}\\
DISPOSALS\_EFFECTIVE\_CSUM\_DIFF &  & && &\textit{0.001}\\
DISPOSALS\_EFFICIENCY\_CSUM\_DIFF \\
FREES\_AGAINST\_CSUM\_DIFF \\
GETS\_GROUNDBALL\_CSUM\_DIFF & &  & && &\textit{0.002} \\
GETS\_GROUNDBALL50\_CSUM\_DIFF & & \textit{0.005} & \textit{0.004}&\textit{0.004}& &\textit{0.007}&\textit{0.005}\\
GOALS\_ACCURACY\_CSUM\_DIFF & &  & && &\textit{-0.021} \\
GOALS\_SHOTS\_CSUM\_DIFF &\textit{0.004}&\textit{0.004}&\textit{0.005}&\textbf{\underline{0.012}}&\textbf{\underline{0.024}}&\textit{0.005}&\textit{0.006}&\textit{0.006}\\
HANDBALLS\_CSUM\_DIFF \\
HITOUTS\_ADVANTAGE\_CSUM\_DIFF \\
HITOUTS\_ADVANTAGE\_RATE\_CSUM\_DIFF  \\
HITOUTS\_WIN\_RATE\_CSUM\_DIFF  \\
INSIDE50\_CSUM\_DIFF &\textbf{\underline{0.004}}&\textit{0.003}&\textit{0.003}&\textit{0.003}&&\textbf{\underline{0.005}}&\textit{0.005}&\textit{0.004} \\
INTERCEPTS\_CSUM\_DIFF  &&\textit{0.002}&\textit{0.002}&&&&\textit{0.003}\\
KICK2HANDBALL\_CSUM\_DIFF&\textbf{\underline{-0.941}}  \\
KICKS\_CSUM\_DIFF \\
KICKS\_EFFECTIVE\_CSUM\_DIFF  &&&&\textit{0.001}&\textit{0.001}\\
KICKS\_EFFICIENCY\_CSUM\_DIFF\\
MARKS\_CONTESTED\_CSUM\_DIFF&&\textit{0.005}&\textit{0.006}\\
MARKS\_CSUM\_DIFF&&&&&\textit{0.001}\\
MARKS\_INSIDE50\_CSUM\_DIFF &&&\textit{0.005}&\textit{-0.008}&\textbf{\underline{-0.015}}&&&\textit{0.008}\\
MARKS\_INTERCEPT\_CSUM\_DIFF \\

MARKS\_ONLEAD\_CSUM\_DIFF\\
METRES\_GAINED\_L4\_CSUM\_DIFF&\textit{0.000}&\textit{0.000}&\textit{0.000}&\textit{0.000}\\
ONE\_PERCENTERS\_CSUM\_DIFF &&&&&&&\textbf{\underline{0.002}}\\
POSSESSIONS\_CONTESTED\_CSUM\_DIFF &&\textit{0.001}&&&&\textit{0.002}&\textit{0.002}\\
POSSESSIONS\_CONTESTED\_RATE\_CSUM\_DIFF &\textit{0.001}\\
POSSESSIONS\_CSUM\_DIFF &\textit{0.000}&&&&&\textit{0.001}\\
POSSESSIONS\_UNCONTESTED\_CSUM\_DIFF&&&&&\textit{0.001}\\
PRESSURE\_CSUM\_DIFF\\
PRESSURE\_DEFENSEHALF\_CSUM\_DIFF& \textbf{\underline{-0.001}}& \textbf{\underline{-0.001}}&\textit{-0.001}\\
REBOUND\_INSIDE50S\_CSUM\_DIFF&\textit{-0.004}&&\textit{-0.004}&\textit{-0.003}&\textit{-0.005}&&\textit{-0.007}\\
SCORE\_LAUNCHES\_CSUM\_DIFF& &\textit{0.004}&\textit{0.006}&\textit{0.009}&\textit{0.009}&\textit{0.004}&\textit{0.007}&\textbf{\underline{0.009}}\\
SPOILS\_CSUM\_DIFF\\
TACKLES\_CSUM\_DIFF&\textit{-0.002}&&&&&\textit{0.003}&\textbf{\underline{0.002}}\\
TACKLES\_INSIDE50\_L4\_CSUM\_DIFF&&\textit{0.005}&&&&\textit{0.007}&\textit{0.006}\\
TURNOVERS\_L4\_CSUM\_DIFF&&&&&\textbf{\underline{0.003}}\\
\bottomrule
    \end{tabular}}
    \end{table}

\begin{table}
    \centering
    \caption{Classification accuracy for the Bradley-Terry team-specific, time-variant expansion, fitted to windows of two seasons and all available data (2015-2023) with both encodings of PIs.}
    \label{tab:accuracy_variantw2}
        \scalebox{0.7}{
    \begin{tabular}{c c c c cc}
    \toprule
    \multicolumn{6}{c}{Window: 2 seasons}\\
    \midrule
    && \multicolumn{2}{c}{Last 4 games cumulative } &\multicolumn{2}{c}{Season cumulative}\\
    \cmidrule(lr){3-4} \cmidrule(lr){5-6}
 Train Season & Test Season & Train Accuracy & Test Accuracy & Train Accuracy & Test Accuracy \\
 \cmidrule(lr){1-2}    \cmidrule(lr){3-4} \cmidrule(lr){5-6}
    2015-16 & 2017 & 69.83\% & 61.27\% & 70.32\% & 60.29\% \\
    2016-17 & 2018 & 67.64\% & 65.05\% & 67.64\% & 65.05\% \\
    2017-18 & 2019 &  66.59\% & 63.29\% & 65.61\% & 63.77\% \\
    2018-19 & 2020 & 68.28\% & 68.12\% & 68.28\% & 66.88\% \\
    2019-20 & 2021 & 68.39\% & 58.82\% & 70.84\% & 54.90\% \\
    2020-21 & 2022 & 65.93\% & 67.48\% & 64.84\% & 67.96\%\\
    2021-22 & 2023 & 68.29\% & 63.55\% & 68.54\% & 64.02\% \\
    2022-23 &  & 70.00\%   &             & 69.05\% & \\
    \midrule
    \multicolumn{6}{c}{All available data}\\
    \midrule
    && \multicolumn{2}{c}{Last 4 games cumulative } &\multicolumn{2}{c}{Season cumulative}\\
    \cmidrule(lr){3-4} \cmidrule(lr){5-6}
 Train Season & Test Season & Train Accuracy & Test Accuracy & Train Accuracy & Test Accuracy \\
 \cmidrule(lr){1-2}    \cmidrule(lr){3-4} \cmidrule(lr){5-6}
    2015-23 &  & 68.54\%   &             & 68.54\% & \\
    \bottomrule
    \end{tabular}}
\end{table}

\begin{table}[H]
    \centering
    \caption{Number of correctly predicted games in the Finals Series for the Bradley-Terry team-specific, time-variant (TS-TV) expansion, fitted to windows of one season (2015-2023) with both encodings of PIs. The Finals Series has 9 games in total.}
    \label{tab:finalsseries_variant_one}
        \scalebox{0.6}{
    \begin{tabular}{c c c c cccc}
    \toprule
    && &\multicolumn{5}{c}{Last 4 games cumulative }\\
    \cmidrule(lr){4-8} 
    Train Season & Test Season & Contest-Specific & TS-TV & Addition & Substitution & Incremental & Majority Voting\\
   \cmidrule(lr){1-3} \cmidrule(lr){4-8}
    2015 & 2016 & 4 & 3 & 3 & 3 & 3 & 3\\
    2016 & 2017 & 6 & 5 & 5 & 5 & 5 & 5\\
    2017 & 2018 & 4 & 4 & 6 & 5 & 6 & 5\\
    2018 & 2019 & 7 & 6 & 6 & 6 & 6 & 6\\
    2019 & 2020 & 4 & 5 & 5 & 3 & 5 & 5\\
    2020 & 2021 & 5 & 3 & 4 & 5 & 5 & 5\\
    2021 & 2022 & 5 & 4 & 6 & 6 & 6 & 6\\
    2022 & 2023 & 5 & 5 & 7 & 6 & 3 & 5\\
\midrule
&& &\multicolumn{5}{c}{Season cumulative }\\
    \cmidrule(lr){4-8} 
    Train Season & Test Season & Contest-Specific & TS-TV & Addition & Substitution & Incremental & Majority Voting\\
   \cmidrule(lr){1-3} \cmidrule(lr){4-8}
    2015 & 2016 & 4 & 3 & 3 & 3 & 2 & 1\\
    2016 & 2017 & 6 & 5 & 4 & 6 & 5 & 5\\
    2017 & 2018 & 4 & 3 & 5 & 6 & 6 & 6\\
    2018 & 2019 & 7 & 7 & 3 & 4 & 5 & 5\\
    2019 & 2020 & 4 & 5 & 5 & 5 & 6 & 4\\
    2020 & 2021 & 5 & 5 & 5 & 5 & 5 & 5\\
    2021 & 2022 & 5 & 3 & 7 & 7 & 6 & 6\\
    2022 & 2023 & 5 & 5 & 4 & 5 & 4 & 4\\
    \bottomrule
    \end{tabular}}
\end{table}

\begin{table}[H]
    \centering
    \caption{Average Accuracy from Experiment 4 Predictions, by Home Team}
    \label{tab:accuracyteams}
    \scalebox{0.7}{
    \begin{tabular}{l *{8}{c}}
    \toprule
    \multicolumn{1}{c}{\textbf{Team}} & \multicolumn{8}{c}{\textbf{Season}} \\
    \cmidrule(lr){2-9}
    & 2016 & 2017 & 2018 & 2019 & 2020 & 2021 & 2022 & 2023 \\
    \midrule
    Adelaide & 79.17\% & 65.48\% & 65.15\% & 65.15\% & 50.00\% & 60.61\% & 72.73\% & 68.06\%\\
    Brisbane Lions & 81.82\% & 77.27\% & 78.79\% & 55.13\% & 87.88\% & 72.22\% & 65.28\% & 89.29\%\\
    Carlton  & 68.18\% & 57.58\% & 66.67\% & 69.70\% & 43.75\% & 63.64\% & 39.39\% & 48.61\%\\
    Collingwood & 40.91\% & 50.00\% & 76.39\% & 66.67\% & 80.95\% & 30.30\% & 47.22\% & 65.56\%\\
    Essendon & 71.21\% & 39.39\% & 51.52\% & 45.45\% & 64.58\% & 57.58\% & 65.15\% & 72.22\%\\
    Fremantle & 59.09\% & 77.27\% & 51.52\% & 46.97\% & 51.67\% & 72.73\% & 65.28\% & 34.72\%\\
    Gold Coast & 78.79\% & 40.91\% & 66.67\% & 84.85\% & 61.11\% & 72.73\% & 59.09\% & 65.15\%\\
    Geelong & 60.26\% & 58.97\% & 69.70\% & 67.95\% & 63.33\% & 79.17\% & 63.10\% & 63.89\%\\
    Greater Western Sydney & 43.06\% & 87.88\% & 66.67\% & 48.61\% & 64.58\% & 54.55\% & 69.70\% & 66.67\%\\
    Hawthorn & 81.94\% & 50.00\% & 58.33\% & 62.12\% & 59.52\% & 51.52\% & 62.12\% & 46.97\%\\
    Melbourne & 60.61\% & 62.12\% & 51.39\% & 40.91\% & 64.81\% & 41.03\% & 47.44\% & 62.50\%\\
    North Melbourne & 77.27\% & 33.33\% & 60.61\% & 69.70\% & 62.50\% & 85.00\% & 72.73\% & 80.30\%\\
    Port Adelaide & 63.64\% & 63.89\% & 59.09\% & 65.15\% & 75.76\% & 66.67\% & 59.09\% & 78.21\%\\
    Richmond & 60.61\% & 68.06\% & 92.31\% & 82.05\% & 71.67\% & 51.67\% & 65.00\% & 53.03\%\\
    Saint Kilda & 69.70\% & 56.06\% & 63.33\% & 69.70\% & 27.78\% & 63.64\% & 63.64\% & 52.78\%\\
    Sydney & 70.24\% & 68.06\% & 48.61\% & 62.12\% & 39.58\% & 58.33\% & 63.89\% & 51.67\%\\
    Western Bulldogs & 59.09\% & 63.64\% & 71.21\% & 42.42\% & 47.92\% & 62.50\% & 51.52\% & 60.61\%\\
    West Coast & 81.94\% & 59.09\% & 60.71\% & 73.61\% & 85.00\% & 54.55\% & 80.30\% & 72.73\%\\
    \bottomrule
    \end{tabular}
    }
\end{table}

\end{document}